\documentclass[final,3p]{elsarticle}

\usepackage{amssymb}

\usepackage{bm}
\usepackage{amsmath}
\newcommand{\bs}{\bm{\xi}^{\rm add}} 
\newcommand{\br}{\mathbf{r}} 
\newcommand{\bp}{\mathbf{p}} 
 
\newcommand{\bq}{\mathbf{q}} 
\newcommand{\bl}{\bm{\xi}}

\newcommand{\cH}{{\cal H}} 
 
\newcommand{\rd}{{\rm d}} 
\newcommand{\cE}{{\cal V}} 
\newcommand{\cF}{{\cal F}}

\journal{Journal of Computational Physics} 
 
\begin{document} 
\begin{frontmatter} 
 
\title{Free energy reconstruction from steered dynamics without post-processing} 
 
\author[aut1,aut2]{Manuel Ath\`enes} 
\author[aut1]{Mihai-Cosmin Marinica}

\address[aut1]{Service de Recherches de M\'etallurgie Physique, D\'epartement des Mat\'eriaux pour le Nucl\'eaire, CEA Saclay, F-91191 Gif-sur-Yvette, France.} 
\address[aut2]{Condensed Matter and Materials Division, Physics and Life Sciences Directorate, LLNL, Livermore, California 94551, USA.}

\begin{abstract} 
Various methods achieving importance sampling in ensembles of nonequilibrium trajectories enable one to estimate free energy differences and, by maximum-likelihood post-processing, to reconstruct free energy landscapes. Here, based on Bayes theorem, we propose a more direct method in which a posterior likelihood function is used both to construct the steered dynamics and to infer the contribution to equilibrium of all the sampled states. The method is implemented with two steering schedules. First, using non-autonomous steering, we calculate the migration barrier of the vacancy in Fe-$\alpha$. Second, using an autonomous scheduling related to metadynamics and equivalent to temperature-accelerated molecular dynamics, we accurately reconstruct the two-dimensional free energy landscape of the 38-atom Lennard-Jones cluster as a function of an orientational bond-order parameter and energy, down to the solid-solid structural transition temperature of the cluster and without maximum-likelihood post-processing. 
\end{abstract}


\end{frontmatter} 

\section{Introduction} 

One important application of molecular simulation is the estimation of 
the Landau free energy $F$ of a given multi-particle system with 
respect to an order parameter $\bl$ 
\begin{eqnarray} 
F (\bl) = - k_BT \ln P(\bl).   
\end{eqnarray} 
where $T$, $k_B$ and $P(\bl)$ denote temperature, Boltzmann's constant 
and the probability to observe the system with value $\bl$ for the 
order parameter, respectively.  Calculating Landau free energies thus 
amounts to measuring occurrence probabilities, a task that molecular 
simulation fails to achieve as soon as relevant portions of the phase 
space are rarely explored. So as to restore numerical ergodicity, many 
simulation techniques have been devised, based on umbrella 
sampling~\cite{torrie:1977}. The generic idea of this technique 
consists in resorting to a judicious steering or restraining potential 
that enhances exploration of regions of phase space that would be 
poorly sampled otherwise. In its usual implementation, a series of 
umbrella sampling simulations~\cite{torrie:1977} are first performed 
so as to cover the various regions of interest, and then the 
collected averages are combined using one of the various reweighing 
procedures~\cite{ferrenberg:1989,kumar:1992,chodera:2008} related to Bennett's 
acceptance ratio method~\cite{bennett:1976} and based on likelihood maximization~\cite{shirts:2003}. 

In this context, Hummer and Szabo~\cite{hummer:2001} proposed to reconstruct the free energy profiles by applying the histogram reweighing procedure to nonequilibrium simulations~\cite{jarzynski:1997} instead of equilibrium simulations. To achieve this, they introduce an additional variable $\xi^{\rm add}$ and connect it to the relevant order parameter $\xi$ via the potential of umbrella sampling. Then, they mechanically steer the additional variable so as to push the particle system along the direction of the order parameter. They finally reconstruct the equilibrium properties by means of a two-step procedure. 
The first step provides the contribution to equilibrium at a given time-slice in trajectory space (after the collected nonequilibrium data have been reweighted using the probability ratios of the reverse-to-forward dynamics~\cite{kurchan:1998,Crooks:1998} within a path-average~\cite{crooks:2000}). In a second step, contributions arising from the entire range of times are finally combined using the weighted histogram analysis method~\cite{ferrenberg:1989,kumar:1992}, as in conventional umbrella sampling. 

We herein propose an estimator enabling one to retrieve equilibrium information from  nonequilibrium trajectories. Like the  aforementioned approaches based on Bennett's acceptance ratio method, our estimator resorts to  reverse-to-forward probability ratios and retrieves information included in all time-slices. 
The estimator will have two advantages : (i) it does not involve any post-processing; (ii) it can be used with more general steering schedules than the one considered by Hummer and Szabo. We will illustrate these two points on the reconstruction of free energy landscapes. 

Concerning point (ii), we will consider Langevin dynamics in which steering arises from additional restraining variables evolving stochastically and autonomously out of equilibrium into otherwise unexplored regions of phase space, enabling enhanced sampling along the steering directions. This way of proceeding can possibly be achieved by coupling the additional variables to high-temperature thermostats~\cite{maragliano:2006,maragliano:2009,abrams:2010}, as in multi-temperature sampling techniques~\cite{Rosso:2001,Rothlisberger:2002}, or by means of an adaptive biasing potential~\cite{laio:2002}. The former approach has been called temperature-accelerated  molecular dynamics (TAMD) and the latter one metadynamics. 


The article is organized as follows. Section~\ref{nonequilibrium} establishes the general theoretical framework for the steered dynamics : the particle system and its additional variables are defined in subsection~\ref{definition}, the equations of the dynamics themselves are introduced in subsection~\ref{steered}, while the reverse-to-forward probability ratios associated with the dynamics, derived in subsection~\ref{derivation}, are used to discuss the two steering schedules for the dynamics in subsection~\ref{autonomous}. In this framework, both the autonomous steering schedule of TAMD and the usual schedules that let a single steering variable evolve non-autonomously at constant speed~\cite{hummer:2001} appear as two particular limiting regimes. The reverse-to-forward probability ratios of subsection~\ref{derivation} are used to construct the two-state estimators~\cite{athenes:2002a,athenes:2002b,athenes:2004,adjanor:2005,Oberhofer:2005,athenes:2007} reviewed in Section~\ref{retrieval} and developed for calculating free energy differences with non-autonomous steering. 
Building on the approaches of Section~\ref{retrieval}, we propose in Section~\ref{multistate} an extended sampler and estimator enabling free energy reconstruction. The derivation that is given works both for autonomous and non-autonomous steering schedules. For completeness, we eventually show how the residence weight algorithm that is proposed relates to the waste-recycling algorithm~\cite{frenkel:2004,ceperley:1977}. 

Section~\ref{application1D} illustrates the performance of the proposed estimator with non-autonomous steering for a one-dimensional reconstruction problem. We compute the migration free energy of a vacancy in Iron ($\alpha$-Fe). The second application given in  Section~\ref{application2D} uses autonomous steering and reconstructs a two-dimensional free-energy landscape of the Lennard-Jones cluster with 38 atoms (LJ$_{38}$). This benchmark system presents a rugged energy landscape, with two energy funnels separated by a high free energy barrier. It has been extensively studied using various methods, which eventually permits one to assess the relative numerical performance of our method with respect to existing methods. Concluding remarks are finally given in Section~\ref{concluding}. 

\section{Extended Hamiltonian and steered dynamics~\label{nonequilibrium}}

\subsection{Extended system~\label{definition}}

Denote by $\br$ the particle position vector of dimension $3I$ 
and by $\bs$ an auxiliary vector of dimension $J$. The potential energy of the particle is $E(\br)$, while 
the steering potential is~\cite{maragliano:2006,laio:2002} 
\[V(\br,\bs) = \frac{1}{2}\sum_{j=1}^{J} \kappa_j |\xi_j^{\rm add}-\xi_j(\br)|^2 \] 
where the order parameter $\bl= (\xi_1,...,\xi_J)$ of  dimension $J$ is represented by a collective variable that is function of the particle positions. We denote the vector positions and vector momenta in the extended system by $\bq$ and $\bp$, respectively. We have $\bq=(\br,\bs)=(r_1,...,r_{3I},{\xi}^{\rm add}_1,...,{\xi}^{\rm add}_{J})$. 
The $i$-th components of these vectors are respectively denoted by $q_i$ and 
$p_i$, with $1 \leq i \leq 3I+J$. Let $m_i$ be the mass associated to the $i$-th 
component, and denote its momentum by $p_i = m_i \dot{q}_i$ where dots above coordinates designate time derivation. 
Denoting $\cE(\bq) = E(\br) + V(\br,\bs)$ the (total) potential energy 
of the extended system and $ \cH (\bp,\bq) = \sum_{i=1}^{3I+J} 
\frac{{p}_i^2}{2m_i} + 
\cE(\bq)$ its Hamiltonian, the normalized canonical 
probability density at temperature $\beta^{-1}=k_BT$ is 
\begin{eqnarray} 
\rho(\bp,\bq) = \frac{1}{h^{3I} I!}{\rm e}^{\beta \cF - \beta \cH(\bp,\bq)  }   \nonumber 
\end{eqnarray}
where the normalizing factor ${\cal F}$ is the Helmholtz free energy of the 
extended system. It is related to the partition function logarithm 
\begin{eqnarray} 
{\cal F} = -\beta^{-1} \ln \left[ \frac{1}{h^{3I}I!}\int{\rm e}^{-\beta 
\cH(\bp,\bq)} {\rm d}\bp {\rm d}\bq \right] .  \nonumber 
\end{eqnarray} 
Here, the infinitesimal volume with respect to coordinates reads 
\begin{eqnarray}
{\rm d}\bp {\rm d}\bq =\prod_{i=1}^{3I+J} {\rm d}p_i {\rm d}q_i .  
\end{eqnarray}
Canonical averages of any quantity $A(\br)$ defined with respect to particle positions can be taken in the extended ensemble ($\omega$ denoting its phase space) as follows  
\begin{eqnarray}
\left\langle A \right\rangle & = & \frac{\int A(\br) \exp\left[-\beta E (\br) \right] {\rm d}\br} {\int \exp\left[-\beta E (\br) \right] {\rm d}\br} \label{numedeno}\\ 
& = & \frac{\int_\omega A(\br) \exp\left[-\beta \cH (\bp,\bq) \right] {\rm d}\bp {\rm d}\bq} {\int_\omega \exp\left[-\beta \cH (\bp,\bq) \right] {\rm d} \bp {\rm d} \bq}  \nonumber  \\ 
& = & \int_\omega A(\br)\rho(\bp,\bq){\rm d}\bp {\rm d}\bq \label{canonical}
\end{eqnarray}
because contributions arising from the additional variables $\bs$ can be inserted inside both integrals in Eq.~\ref{numedeno}. Hence, the additional variables and potentials do not affect the thermodynamic expectations of the particle system. 

\subsection{Steered Langevin dynamics~\label{steered}} 

We consider that any coordinate $q_i$ with $1 \leq i \leq 3I+J$ 
is coupled to an independent thermal reservoir at temperature 
$T$. The traditional Langevin dynamics amount to propagating the system by 
solving the equations of motion ($f_i = -\partial_{q_i} \cH$) 
\begin{eqnarray} 
\dot{q}_i  = m_i^{-1} p_i & \hspace{2cm}&
\dot{p}_i  = f_i - \gamma_i p_i + b_i(t) \sqrt{2\gamma_i k_B T m_i}\hspace{1.5cm} \label{position} 
\end{eqnarray} 
where $b_i$ represents a white noise of amplitude 1 and zero 
mean, while $\gamma_i$ denotes the friction characterizing the coupling 
intensity with the $i$-th thermal bath. Here the amplitude of the 
fluctuations $\sqrt{2\gamma_i k_B T m_i}$ determines the temperature 
$T$.

Let us assume that we have prepared the system in thermodynamic equilibrium at time $t=0$ [e.g. by propagating the Langevin dynamics~\eqref{position} long enough and then setting the time to zero]. 
At $t=0$,  we switch on the external forces $f_j^{\rm ext}=(\mu_j-1)f_j$ to act mechanically upon on the additional variables $q_j$. The rescaling factors $\mu_j$ are such that $0 \leq \mu_j  \leq 1$ for  $3I < j \leq 3I+J$. We also define $ \mu_i=1 $ for $1 \leq i \leq 3I $ by extension and we have  $f_j^{\rm ext}+f_j=\mu_jf_j$. The frictional forces and the square of the fluctuations acting upon the additional variables are rescaled in the same way using the $\mu_j$'s. The extended system is then propagated for a duration $\tau$ using the steered Langevin dynamics below (using the convention on the indices, $1 \leq i \leq 3I$ and $3I< j \leq 3I+J$) : 
\begin{eqnarray} 
\dot{q}_i =  m_i^{-1} p_i  & \hspace{2cm} &
\dot{p}_i =  f_i - \gamma_i p_i + b_i(t) 
\sqrt{2\gamma_i k_B T m_i} \\
\dot{q}_j =  m_j^{-1} p_j  & \hspace{2cm} &
\dot{p}_j =  \mu_j {f_j} -\mu_j \gamma_j p_j + b_j(t) \sqrt{2\mu_j \gamma_j 
k_BT m_j} \label{gs} . 
\end{eqnarray}
The rescaling of the dynamics amounts to unbalancing the interactions between particles and additional variables and to decreasing the coupling 
intensity with the  $j$-th thermal bath while maintaining constant its temperature $T$. 
The effect of the external forces is to make the system depart from thermodynamic equilibrium, by reducing the restraining forces 
acting upon the additional variables, which enhances exploration of phase space along the variables $\bs$. Additional commentaries about the 
dynamics~\eqref{gs} have been deffered to subsection~\ref{autonomous} because they are based on the reverse-to-forward probability ratio derived in 
Section~\ref{derivation}. This ratio will indeed quantify the deviation of the dynamics with respect to equilibrium and will enable one to construct the estimators of Sections~\ref{retrieval} and~\ref{multistate}. 

\subsection{Reverse-to-forward probability ratio and discretization~\label{derivation}}

The probability to generate the dynamics along a path $z=\left[\bq(t)\right]_{0 \leq t \leq \tau}$ given that the system is at $\left( \bq(0),{\bp}(0) \right)$ at time $t=0$ is denoted  $\varrho_F(z)$, while the probability to generate the same path using the reverse dynamics starting from system $ \left( \bq(\tau),{\bp}(\tau)\right) $ at time $t=\tau$ down to $t=0$ is denoted  $\varrho_R(z)$. 
Then, the reverse-to-forward probability ratio and the work ${\cal W } (z)$ done by the external forces $f^{\rm ext}_{3I < j \leq 3I+J}$ upon the extended system are related by the following expression (Eq. 2 in~\cite{jarzynski:2008})
\begin{eqnarray}
\frac{\rho\left(\bp(\tau),\bq(\tau)\right)\varrho_R(z)}{\rho\left(\bp(0),\bq(0) \right)
\varrho_F(z)} & = & \exp\left[-\beta {\cal W } (z) \right] \label{BD1}. 
\end{eqnarray}

The identity above and the expression of the work will be derived explicitly for the discretized Langevin dynamics in this subsection. Before, we point out that identity~\eqref{BD1} is similar to the 
more well-known identity involving the reverse-to-forward probability ratio due to Crooks~\cite{Crooks:1998}, except that a difference of free energy between a target system and a reference system appears in the latter form. No free energy difference appears here because the target and reference systems are the extended system itself with the extended Hamiltonian. The thermodynamic implications involving the two mentioned identities are compared in Ref.~\cite{jarzynski:2008}. Besides, from a mathematical perspective, Eq.~\ref{BD1} can be interpreted as a generalized detailed balance equation involving the forward and backward Kolmogorov operators associated to our Langevin dynamics (see Eq. 4.43 in Ref.~\cite{LRS2010}). This interpretation allows both to define time reversibility rigorously and to extend the original derivations~\cite{Crooks:1998,jarzynski:2008}, which considered discrete-time Markov processes, to general continuous-time Langevin dynamics such as the one considered here. 

Since in practical applications we have to discretize the dynamics, we are authorized to expand the reverse and forward conditional probabilities, so as
to include these quantities in the estimators directly. This is the approach that we follow in the sequel.  
Let $\Delta t$ denote the discretization time step and  $\chi_n$ denote a state $\left(\bq(t_n),\bp(t_n)\right)$ at time $t_n=n \Delta t$. 
The discretized trajectory of a path $z$ is characterized by the successive states  $(\chi_0,...,\chi_n,...,\chi_N)$ obtained at times $(t_0,...,t_n,..., t_N)$ by propagating the Langevin dynamics {\em forward} starting from a given state $\chi_0$.  This is achieved by updating the following discretization  scheme~\cite{athenes:2004,adjanor:2005,athenes:2006,athenes:2008,bussi:2009} from time $t_{0}$ to time $t_N$ ($1\leq i \leq 3I+J$) 
\begin{subequations}
\begin{align}
{p}_{i,k+1/4}   & =  {p}_{i,k} e^{-\tilde{\gamma}_i \Delta t/2}+ {\eta}^+_{i,k+1/{4}} \label{forward0}\\ 
{p}_{i,k+1/{2}} & =  {p}_{i,k+1/{4}}+\tilde{f}_{i,k}\Delta t/2 \label{forward1} \\ 
{q}_{i,k+1}     & =  {q}_{i,k} + {p}_{i,k+1/{2}} \Delta t/m_i \label{forward2} \\ 
{p}_{i,k+3/{4}} & =  {p}_{i,k+1/{2}} +\tilde{f}_{i,k+1} \Delta t/2  \label{forward3} \\ 
{p}_{i,k+1}     & =  {p}_{i,k+3/{4}}  e^{-\tilde{\gamma}_i \Delta t/2} +   {\eta}^+_{i,k+3/{4}}   \label{forward4}
\end{align}
\end{subequations} 
where index $k$ denotes time $k\Delta t$ while $\tilde{f}_{i,k}=\mu_i{f}_i(k\Delta t)$ and $\tilde{\gamma}_{i}=\mu_i\gamma_i$  ($\mu_i=1$ if $i\leq 3I$). Besides, the noises ${ \eta}^+_{i,k+1/{2}\pm 1/{4}}$ in Eqn.~\eqref{forward0} and~\eqref{forward4} are normal and have mean zero and variance $\sigma_i=(1-{\rm e}^{-\tilde{\gamma}_i \Delta t})m_i/\beta$. 
Updates~\eqref{forward0} and~\eqref{forward4} correspond to the momentum variations due to two consecutive Ornstein-Uhlenbeck processes of duration $\frac{\Delta t}{2}$.  
These processes consist of propagating the momentum $p_i$ using 
\begin{equation} 
\dot{p}_i=-\tilde{\gamma}_i p_i + b_i(t) \sqrt{2 m_i \tilde{\gamma}_i/\beta}  \label{o-u}
\end{equation}
from $t=t_n$ to $t_n+\Delta t /2$ and from $t=t_n+\Delta t/2$ to $t_{n+1}$, where $b_i(t)$ is an uncorrelated white noise of unit amplitude. 

Because the scheme corresponds to a double Strang-Trotter decomposition~\cite{adjanor:2005} with the position update in the center and then half momentum updates with respect to the force and the stochastic processes, it is symmetric and can thus be updated or downdated depending on whether the dynamics is considered to be forward or reverse. For the reverse dynamics, we must iterate 
\begin{subequations}
\begin{align}
{p}_{i,k+3/{4}} & =  {p}_{i,k+1}  e^{-\tilde{\gamma}_i \Delta t/2} + {\eta}^-_{i,k+3/{4}}   \label{reverse4} \\
{p}_{i,k+1/{2}} & =  {p}_{i,k+3/{4}} -\tilde{f}_{i,k+1} \Delta t/2  \label{reverse3} \\ 
{q}_{i,k}       & =  {q}_{i,k+1} - {p}_{i,k+1/{2}} \Delta t/m_i \label{reverse2} \\ 
{p}_{i,k+1/{4}} & =  {p}_{i,k+1/{2}}  -\tilde{f}_{i,k}\Delta t/2 \label{reverse1} \\ 
{p}_{i,k}       & =  {p}_{i,k+1/{4}} e^{-\tilde{\gamma}_i \Delta t/2}+{\eta}^-_{i,k+1/{4}} \label{reverse0}
\end{align}
\end{subequations}
where the reverse noises ${ \eta}^-_{i,k+1/{2}\pm 1/{4}}$ have the same variance. Besides, the time reversal of the  Ornstein-Uhlenbeck process in~\eqref{reverse4} and~\eqref{reverse0} is the process itself. 

We then denote the probabilities of the discretized dynamics by ${\rm P}_{\rm cond}(z|\chi_N,N)$ and ${\rm P}_{\rm cond}(z|\chi_0,0)$.  They will approximate the quantities $\varrho_R(z)$ and $\varrho_F(z)$ in~\eqref{heat}. 
As a result of the discretization, the forward path probability can be factorized into the following product 
\begin{eqnarray}
{\rm P}_{\rm cond }(z|\chi_0,0) &= &\prod_{i=1}^{3I+J} \prod_{k=0}^{N-1} {\Phi}_{\sigma_i}(\eta^+_{i,k+1/{4}}) {\Phi}_{\sigma_i}(\eta^+_{i,k+3/{4}}) \\
&= &\prod_{i=1}^{3I+J} \prod_{k=0}^{N-1} A^2_{\sigma_i} \exp \left\lbrace -\frac{(2m_i \beta)^{-1}}{1-e^{-\tilde{\gamma}_i \Delta t}}\left[ (\eta^+_{i,k+1/{4}})^2 + (\eta^+_{i,k+3/{4}})^2 \right] \right\rbrace . 
\end{eqnarray}
where ${\Phi}_{\sigma_i}$ stands for the normal probability of variance  $\sigma_i=m_i(1-e^{-\tilde{\gamma}_i \Delta t})/\beta$ and $A_{\sigma_i}$ denotes its normalizing factor. The normal laws ${\Phi}_{\sigma_i}$ are used to generate the 
stochastic noises $\eta^+_{i,k+1/{4}}$ and $\eta^+_{i,k+3/{4}}$ of the $i$-th thermostat along trajectory $z$ ($0 \leq k < N$). The conditional probability to generate $z$ backward can be  
decomposed into a similar product of normal probabilities 
\begin{eqnarray}
 &= &\prod_{i=1}^{3I+J} \prod_{k=0}^{N-1} {\Phi}_{\sigma_i}(\eta^-_{i,k+1/{4}}) {\Phi}_{\sigma_i}(\eta^-_{i,k+3/{4}}) \\
&= &\prod_{i=1}^{3I+J} \prod_{k=0}^{N-1} A^2_{\sigma_i} \exp \left\lbrace -\frac{(2m_i \beta)^{-1}}{1-e^{-\tilde{\gamma}_i \Delta t}}\left[ (\eta^-_{i,k+1/{4}})^2 +  (\eta^-_{i,k+3/{4}})^2 \right] \right\rbrace. 
\end{eqnarray}

Let $Q_{i,k}$ denote the temperature-scaled logarithm of the reverse-to-forward probability ratio associated with the Ornstein-Uhlenbeck processes for the $i$-th thermostat at step $k$. We have 
\begin{eqnarray}
Q_{i,k} & = &  \beta^{-1} \left\lbrace \ln \left[ {\Phi}_{\sigma_i}(\eta^-_{i,k+1/{4}}) {\Phi}_{\sigma_i}(\eta^-_{i,k+3/{4}}) \right] - \ln \left[ {\Phi}_{\sigma_i}(\eta^+_{i,k+1/{4}}) {\Phi}_{\sigma_i}(\eta^+_{i,k+3/{4}}) \right] \right\rbrace  \label{chaleur1} \\
& = & \frac{(2m_i)^{-1}}{1-e^{-\tilde{\gamma}_i \Delta t}}  \left\lbrace \left[ (\eta^+_{i,k+1/{4}})^2 + (\eta^+_{i,k+3/{4}})^2 \right] -\left[ (\eta^-_{i,k+1/{4}})^2 + (\eta^-_{i,k+3/{4}})^2 \right] \right \rbrace \label{chaleur2}
\\ 
& = & \frac{1}{2m_i}  \left\lbrace p^2_{i,k+1} - p^2_{i,k+3/{4}} + 
p^2_{i,k+1/{4}} -p^2_{i,k} \right\rbrace \label{chaleur3} \\ 
& = & \frac{1}{2m_i} \left[ p^2_{i,k+1}-p^2_{i,k}  \right] + \frac{\Delta t^2}{8m_i} \left[ \tilde{f}^2_{i,k+1}-\tilde{f}^2_{i,k}  \right]
-\frac{1}{2} (q_{i,k+1}-q_{i,k})\cdot(\tilde{f}_{i,k+1}+\tilde{f}_{i,k}) \label{chaleur4}. 
\end{eqnarray}
The transformation from~\eqref{chaleur2} to~\eqref{chaleur3} involves expressing the noises as a function of the momenta after and before the Ornstein-Uhlenbeck processes and yields a form of detailed balance. In the transformation from~\eqref{chaleur3} to ~\eqref{chaleur4}, the intermediate momenta $p_{i,k+1/{4}}$ and $p_{i,k+3/{4}}$ have been expressed as a function of the forces and positions at integer steps. 

The effective work done along the path from $t_0$ to $t_n=\frac{n}{N}\tau$ defined by~\cite{athenes:2004}
\begin{eqnarray} 
W_n & = & -\beta^{-1} \left[ \ln \frac{ \rho(\chi_n) {\rm P}_{\rm cond }(z|\chi_n,n) } {\rho(\chi_0) {\rm P}_{\rm cond }(z|\chi_0,0) }\right] \nonumber \\
   & = & \cH(\chi_n)-\cH(\chi_{0}) - \sum_{i=1}^{3I+J}\sum_{k=0}^{n-1} Q_{i,k}  \label{effectivework2}.  
\end{eqnarray} 
can be evaluated from the knowledge of the trajectory via the $Q_{i,k}$'s. 
The effective works will be used to retrieve equilibrium information in Section~\ref{multistate}. Nevertheless, from a thermodynamical point of view, it is instructive to formulate the work ${\cal W}(z)$ in the continuum limit, achieved here when $N$ goes to infinity and with $\Delta t = \tau/N$. We thus define ${\cal Q}_i(z)$ as 
\begin{eqnarray} 
{\cal Q}_i(z) & = & \lim_{N\rightarrow + \infty} \sum_{k=0}^{N-1} Q_{i,k}  . 
\end{eqnarray} 
with $\Delta t = \tau/N$ and all states $\chi(t_k)$ inside the continuous path $z$ when defining the limit. 
From the relation~\cite{athenes:2004} 
\begin{eqnarray}
 \sum_{k=0}^{N-1} Q_{i,k} & = &  \frac{1}{2m_i} \left[ p^2_{i,N}-p^2_{i,0}  \right] + \frac{\Delta t^2}{8m_i} \left[ \tilde{f}^2_{i,N}-\tilde{f}^2_{i,0}  \right]
-\frac{1}{2} \sum_{k=0}^{N-1} (q_{i,k+1}-q_{i,k})\cdot(\tilde{f}_{i,k+1}+\tilde{f}_{i,k}) \nonumber,   
\end{eqnarray}
we deduce 
\begin{eqnarray}
{\cal Q}_i(z)  & = & \frac{1}{2m_i} \left[ p^2_{i}(\tau)-p^2_{i}(0)\right] - \int_0^\tau \frac{\rd q_i}{\rd t} \mu_i f_i {\rm d}t \label{travail},  
\end{eqnarray} 
recalling that $\mu_i f_i = \tilde{f}_i$. Neglecting the constant Ito term arising from the integration of $p_idp_i$, 
the quantity ${\cal Q}_i(z)$ can be interpreted as the work 
done along the trajectory $z$ by the force 
\[\ell_i = \frac{\rd {p}_i} {\rd t}- \mu_if_i 
=-\tilde{\gamma}_i {p}_i+\sqrt{2\tilde{\gamma}_ik_B T m_i}b_i(t)\] 
that is exerted by the $i$-th thermostat upon the $i$-th coordinate ($\tilde{\gamma_i} = \mu_i\gamma_i$). 
This quantity thus represents the heat exchanged with the $i$-th thermostat. We recover an additional result in the continuum limit : the total heat exchanged with the thermostats during the forward dynamics, defined by 
\begin{eqnarray}
 {\cal Q}(z) = \sum_{i=1}^{3I+J} {\cal Q}_i(z)  \label{totalheat}
\end{eqnarray} 
relates to the ratio of the reverse-to-forward conditional probability via the well-known expression~\cite{Crooks:1998,crooks:2000}
\begin{eqnarray}
\frac{\varrho_R(z)}{\varrho_F(z)} & = & \exp\left[\beta {\cal Q } (z) \right]. \label{heat}
\end{eqnarray}
The heat also relates to the quantity ${\cal W}(z)$ defined in~\eqref{BD1} via a conservation equation~\eqref{conservation}, obtained by inserting~\eqref{heat} into~\eqref{BD1} and then resorting to the relations $\rho(\bp,\bq) = \exp \left[\beta \left({\cal F }-{\cal H} (\bp,\bq) \right) \right]$ both at $t=0$ and $t=\tau$. We have \begin{equation}
{\cal Q } (z) = \cH \left(\bp(\tau),\bq(\tau)\right) - \cH \left(\bp(0),\bq(0)\right) - {\cal W } (z).  \label{conservation}
\end{equation}
Besides, resorting to $f_i=-\partial_{q_i}\cH$ in Eq.~\ref{travail} and then summing yields an additional relation for the total heat exchanged with the thermostats 
\begin{eqnarray}
{\cal Q}(z) & = & \sum_{i=1}^{3I+J}\frac{1}{2m_i} \left[ p^2_{i}(\tau)-p^2_{i}(0)\right] + \int_{0}^{\tau} \left[\nabla_{\bq}{\cH} 
\cdot \frac{\rd \bq}{\rd t} -\sum_{i=1}^{3I+J}
(1-\mu_i) \partial_{q_i} {\cal H}\frac{\rd q_i}{\rd 
t} \right] \rd t \nonumber \\
& = & \left\lbrace {\cH}\left[\bp(\tau),\bq(\tau)\right]-\cH \left[ \bp(0),\bq(0)\right] \right\rbrace-\sum_{i=3I+1}^{3I+J} \int_{q_i(0)}^{q_i(\tau)} (1-\mu_i) \partial_{q_i} \cH {\rm d} 
q_i(t). ~\label{nonconservation} 
\end{eqnarray} 
Note that the last summation runs from $3I+1$ since  $\mu_i=1$ for $0 \leq i \leq 3I$. Then, substracting~\ref{conservation} to~\ref{nonconservation} enables one to identify ${\cal W } (z)$ explicitly
\begin{eqnarray}
{\cal W } (z) & = &  \sum_{j=3I+1}^{3I+J} \int_{q_j(0)}^{q_j(\tau)} (1-\mu_j) \partial_{q_j} \cH {\rm d}
q_j(t) \label{work}. 
\end{eqnarray}
This quantity indeed corresponds to the work done by the external forces $f_j^{\rm ext}=(1-\mu_j) \partial_{q_j} \cH$ upon the extended system, as stated in~\cite{chernyak:2006}. 

The limiting case consisting of setting $\mu_j=1$ for all $j>3I$ cancels the work in Eq.~\ref{work} [${\cal W}(z)=0$]. This in turn implies a specific form of detailed balance 
$\rho\left[\chi(\tau)\right]\varrho_R(z)=\rho\left[\chi(0)\right]\varrho_F(z)$ 
ensuring that the dynamics sample the equilibrium distribution 
$\rho(\chi)=\propto \exp\left[-\beta \cH \left(\chi\right) \right]$. 
Aside from this limiting case, two particular schedules are possible for the steered dynamics depending on the $\mu_j$-values in Eq.~\ref{work}, which we discuss below. 

\subsection{Autonomous versus non-autonomous steering~\label{autonomous}}

A first steering regime appears when the values of the scaling factors $\mu_j$'s are set to zero for all $j>3I$. The noise amplitude and the friction $\tilde{\gamma}_j = \mu_j \gamma_j$ vanish (see Eq.~\ref{gs}). 
Any coordinate $q_j$ then evolves at a constant imposed velocity as in the schedule established by Hummer and Szabo~\cite{hummer:2001}. The forces are conservative and time-dependent with respect to the real particles (once the additional variables have been eliminated by solving for them). The dynamics is said to be non-autonomous~\cite{chernyak:2006} and we refer to this regime as non-autonomous steering. Non-autonomous dynamics with $J=1$ guided by (time-dependent) conservative forces are well suited for computing free-energy profiles in one dimension or differences of free energy. Note that the fast switching schedule introduced by Jarzynski~\cite{jarzynski:1997} amounts to non-autonomous scheduling with a single external parameter $\lambda(t) \equiv q_{3I+1}(t)$ and $\mu_{3I+1}=0$ in Eq.~\ref{work}. Furthermore, the integral form in Eq.~\ref{work} with $J=1$ corresponds to Jarzynski's definition of the work provided we consider the additional variable as a coupling parameter acting upon the Hamiltonian of the particle subsystem. 

The second steering regime consists of choosing $0<\mu_j<1$ for $j>3I$. In this regime, the extended Langevin dynamics of Section~\ref{steered} is autonomous : the additional variables evolve stochastically by means of a force field $\left\{ \mu_i {f}_i \right\}_{1\leq i \le 3I+J}$ that is time-independent and non-conservative~\cite{chernyak:2006}, i.e. that does not derive from a potential function except for particular conditions on the forces and the $\mu_j$'s. We refer to this regime as autonomous steering. As will be shown in Section~\ref{application2D}, autonomous steering is well adapted to the use of more than one additional variable. 

In the second regime, the dynamics of the additional variables may be given to a different thermodynamic interpretation. Indeed, the additional variables~\eqref{gs} also evolve according to the equation ($\widetilde{m}_j= \mu_j^{-1} m_j$, $j>3I$)
\begin{eqnarray}
\ddot{q}_j = \widetilde{m}_j^{-1} f_j -\tilde{\gamma}_j \dot{q}_j + b_j\sqrt{ 2 \widetilde{m}_j^{-1}\tilde{\gamma}_j k T_j} \nonumber
\end{eqnarray}
where $T_j=\mu_j^{-1}T$ denotes the effective temperature of the thermostat that is actually coupled to $q_j$ and $\widetilde{m}_j$ denotes an effective mass for the additional variable. This dynamics is a particular implementation of the general dynamics given in~\cite[Eq.3]{vanden:2009}. 
Dynamics coupled to thermostats at different temperatures reach a nonequilibrium steady-state with no well-defined temperature and satisfy a generalized detailed balance equation~\cite{bodineau:2007,crookthesis}. Fluctuation theorems as well as reverse-to-forward probability ratios considered with respect to multi-temperature dynamics then relate to the heat transfers between the system and the various thermostats around the nonequilibrium steady state~\cite{bodineau:2007}. 
The rescaling of the forces in the dynamics (subsection~\ref{steered}) actually ensures that the reverse-to-forward path probability ratios relate to a transient mechanical work rigorously defined with respect to the equilibrium distribution of interest, as in the steering protocol of Hummer-Szabo. In particular, the work~\eqref{work} depends on the potential energy of the extended Hamiltonian and not on its kinetic energy. 

Note that the stationary distribution reached by a multi-temperature dynamics exhibits a known analytical form~\cite{Rosso:2001,Rothlisberger:2002,maragliano:2006} when a separation of frequencies occurs between a slow variable $q_j$ subject to a thermostat at high temperature $T_j$ and the remaining fast variables at normal temperature $T$. From this analytical form, the equilibrium probability profile ${\rm P}^{\rm eq}_T$ of $q_j$ at temperature $T$ can be extracted from the established relation $P^{\rm eq}_T(q_j) \propto \left[ {\rm P}^{\rm st}(q_j) \right]^{T_j/T}$ where ${\rm P}^{\rm st}(q_j)$ is the stationary probability profile measured during a simulation. No separation of frequencies needs to be imposed  in the approach of the present paper where the rescaling factor $T_j/T=\mu_j^{-1}$ acts upon the dynamics directly~\eqref{gs}. 

In order to retrieve equilibrium information from transient nonequilibrium dynamics, Jarzynski derived its remarkable identity that involves the exponential average of the work (refer to~\cite{jarzynski:1997,bochkov77,bochkov81,jarzynski:2004,jarzynski:2008,jarzynski:2008b} for original and review papers on fluctuation theorems). We now briefly review the computational extensions that have been made to this approach for non-autonomous steering with a single additional variable. 

\section{Two-state estimators for non-autonomous steering\label{retrieval}~\label{BS}}

We assume here that trajectories are generated using a single steering variable $\lambda$ and with non-autonomous scheduling ($\mu_{3I+1}=0$). The phase space of non-autonomous paths is defined as follows 
\begin{equation}
\Omega_{\rm na} = \left\lbrace z \hspace{0.15cm}{\rm such \hspace{0.15cm} that \hspace{0.15cm} } \forall n,\hspace{0.15cm} q_{3I+1}(t_n) = \lambda_n \right\rbrace. \nonumber
\end{equation}

A simple estimator associated to a biased sampler can give access to the ratio of normalizing constants related to the two thermodynamic states defined by ${\lambda}_0$ and ${\lambda}_N$~\cite{athenes:2002b,athenes:2004,adjanor:2005,Oberhofer:2005,lechner:2007,athenes:2007}.  
This ratio can indeed be cast in the following form 
\begin{eqnarray}
\frac{\int \delta [q_{3I+1}-\lambda_N] \rho(\chi)\rd \chi}{\int \delta [q_{3I+1}-\lambda_0] \rho(\chi)\rd \chi} & = & \frac{\int_{\Omega_{\rm na}} {\rm P}_{\rm cond}(z |\chi_N,N)\rho(\chi_N) {\cal D} z}{\int_{\Omega_{\rm na}}{\rm P}_{\rm cond} (z|\chi_0,0)\rho(\chi_0) {\cal D} z } \label{transformation1}\\
&= &\frac{\int_{\Omega_{\rm na}} \left[ {{\rm P}_{\rm cond}(z|\chi_N,N)\rho(\chi_N)} /{ {\rm P_B}^\varphi(z)} \right] {\rm P_B}^\varphi(z) {\cal D }z}  {\int_{\Omega_{\rm na}} \left[{{\rm P}_{\rm cond}(z|\chi_0,0)\rho(\chi_0)} /{{\rm P_B}^\varphi(z)} \right]{\rm P}_{\rm B}^\varphi(z) {\cal D}z } \label{pathave}. 
\end{eqnarray}
The first transformation~\eqref{transformation1} merely exploits the normalization of conditional probabilities with respect to path space $\Omega_{\rm na}$. 
The second transformation~\eqref{pathave} formally inserts the biased probability distribution ${\rm P}^\varphi$ with respect to which sampling is performed (note that Jarzynski's identity is recovered by replacing ${\rm P}_B^\varphi(z)$ with ${\rm P}_{\rm cond}(z|\chi_0,0)\rho_0(\chi_0)$ where $\rho_0(\chi)$ is the equilibrium density conditioned on $q_{3I+1}=\lambda_0$). 
In applications~\cite{athenes:2004,adjanor:2005,Oberhofer:2005,lechner:2007} of identity~\eqref{pathave}, the biasing potential $\varphi$ is a function of the work function ${\cal W}(z) =-\beta^{-1} \ln\left\{ \left[{\rm P}(z|\chi_N,N)\rho(\chi_N)\right] / \left[ {\rm P}(z|\chi_0,0)\rho(\chi_0) \right] \right\}$. In practice, the residence weight algorithm~\cite{athenes:2002a} was observed to achieve good performance~\cite{athenes:2002b,athenes:2007} (see Appendix~\ref{workbiased}). 

The purpose of the paper is to extend the residence weight algorithm so that its sampler and associated estimator can handle the multiple thermodynamic states that can be defined owing to the extended system, irrespective of whether the scheduling of the steered dynamics is non-autonomous or autonomous. 

\section{Multi-state estimator~\label{multistate}}

Residence weight algorithms can be formulated from two opposite points of view~\cite{athenes:2007}. Here, we first build both the sampler and the estimator of the algorithm upon Bayes theorem by adopting the viewpoint of statistical inference. Then, we reinterpret the estimator as a conditional expectation (second viewpoint) in order to show the connection with the waste-recycling algorithm. 

\subsection{Posterior likelihood viewpoint~\label{rwa}}

Here, a marginal probability will be the importance function with respect to which path sampling is achieved,   
while a posterior likelihood function will be used on-the-fly to infer the equilibrium contribution of each generated state within the estimator. 
The marginal probability is defined in the path ensemble as the  {\it a priori} probability of witnessing a path $z$ under all possible hypotheses, i.e. as the sum of the product of all probabilities of hypotheses ${\rm P}^\varphi$ and corresponding conditional probabilities ${\rm P_{cond}}$:
\begin{equation}
{\rm P}^\varphi_{\rm M}(z) = \sum_{n=0}^{N}{\rm P_{cond}}(z|\chi_{n},n) \cdot {\rm P}^\varphi(\chi_n,n).~\label{marginal}
\end{equation}
An hypothesis $(\chi_n,n)$ is the knowledge of a state belonging to the path and of its index. 
The conditional probabilities in Eq.~\ref{marginal} are given by~\cite{dellago:2002,athenes:2006,stoltz:2007}
\begin{eqnarray}
{\rm P}_{\rm cond }(z|\chi_n,n) &= &\prod_{i=1}^{3I+J} \prod_{k=n}^{N-1} {\Phi}_{\sigma_i}(\eta^+_{i,k+1/{4}}) {\Phi}_{\sigma_i}(\eta^+_{i,k+3/{4}})  \prod_{k=n-1}^{0} {\Phi}_{\sigma_i}(\eta^-_{i,k+1/{4}}) {\Phi}_{\sigma_i}(\eta^-_{i,k+3/{4}}). \label{shooting}
\end{eqnarray}
where the normal distribution ${\Phi}_{\sigma_i}$ are detailed in subsection~\ref{derivation}.  
Here, ${\rm P}_{\rm cond }(z|\chi_n,n)$ is the probability to generate the states $\chi_{n+1},  \chi_{n+2},... \chi_{N}$ starting from any $\chi_{n}$ by updating Eqn.~\ref{forward0}-\ref{forward4} and then to generate the states $\chi_{n-1}, \chi_{n-2},... \chi_{0}$ starting from $\chi_{n}$ by downdating Eqn.~\ref{reverse4}-\ref{reverse0}. The two particular cases ${\rm P}_{\rm cond} (z|\chi_0,0) \approx \rho_F (z)$ and ${\rm P}_{\rm cond} (z|\chi_N,N) \approx \rho_R (z)$ were considered previously in biased path sampling schemes~\cite{athenes:2004,adjanor:2005,Oberhofer:2005,lechner:2007,athenes:2007} to denote the probability to generate the forward and reverse trajectories, respectively. 

Additionally, the prior probability of hypothesis $(\chi,n)$ in Eq.~\ref{marginal} is 
\begin{equation}
{\rm P}^\varphi(\chi,n) = 
\begin{cases} 
\rho^\varphi(\chi) h^{\chi}(n)   & \text{  for non-autonomous scheduling,}\\ 
\rho^\varphi(\chi)\frac{1}{N+1} & \text{  for autonomous scheduling.}
\end{cases} \nonumber
\end{equation} 
For non-autonomous scheduling, $h^\chi$ denotes the prior probability of index $n$ and is such that $h^\chi(n)=1$ if $\chi$ pertains to the sliced phase space 
$\omega_n = \left\{ (\bp,\bq) \hspace{2mm} | \hspace{2mm} q_{3I+1}= \lambda_n \right\}$ that corresponds to index $n$, otherwise $h^\chi(n)=0$. Whatever $\chi$, we also assume that $h^\chi(n)h^\chi(m)=0$ for $n \neq m$ and $\sum_{n=0}^{N+1}h^\chi(n)=1$ : the steering amplitude captures once all important regions of phase space.
With autonomous scheduling, $h^\chi$ reduces to $\frac{1}{N+1}$ whatever $\chi$, involving the independence of the slice index from $\chi$. 
Besides, $\rho^\varphi$  denotes a biased prior distribution of states 
\begin{equation}
\rho^\varphi(\chi)  = \frac{1}{h^II!}\exp\left[\beta \cF^\varphi - \varphi(\chi) -\beta \cH(\chi) \right] 
\end{equation}
where the biasing potential  $\chi \rightarrow \varphi(\chi)$ is here a state function (rather than a work function as in previous implementations) and the normalizing constant $\cF^\varphi$ is the $\varphi$-dependent free energy. Irrespective of the scheduling, we have the useful equality  
\begin{eqnarray}
\rho^\varphi(\chi) & = & \sum_{n=0}^N\int_{\omega} \delta(\chi-\chi_n) {\rm P}^\varphi(\chi_n,n) \rd \chi_n. 
\end{eqnarray}
where $\omega$ denotes the unrestricted phase space (we previously assumed $\omega=\cup_{n=0}^{N+1}\omega_n$ for non-autonomous scheduling). 
This equality will enable us to express the biased density $\rho^\varphi$ as a path integral of marginal and posterior probabilities, irrespective of whether the scheduling is {\em autonomous} or {\em non-autonomous}. For this purpose, we introduce $\Omega(\chi_n,n)$ to denote the subspace of all paths going through $\chi_n$ (at slice index $n$) and exploit the property that the sum over the path probabilities conditioned on $\chi_n$ is normalized to one in $\Omega(\chi_n,n)$ 
\begin{eqnarray}
\rho^\varphi(\chi) & = & \sum_{n=0}^N \int_{\omega}  \left[ \int_{\Omega(\chi_n,n)} {\rm P}_{\rm cond}(z|\chi_n,n){\cal D}z \right]  \delta(\chi-\chi_n) {\rm P}^\varphi(\chi_{n},n) {\rm d} \chi_n \nonumber \\
& = & \sum_{n=0}^N \int_\Omega \delta(\chi-\chi_n) {\rm P}_{\rm cond}(z|\chi_n,n) {\rm P}^\varphi(\chi_n,n) {\cal D}z \label{sumdirac} \\ 
& = & \int_\Omega \left[ \sum_{n=0}^N \delta(\chi-\chi_n) {\rm P}_{\rm sel}(\chi_{n},n|z) \right] {\rm P}^\varphi_{\rm M}(z) {\cal D}z 
\label{sumdirac2}
\end{eqnarray}
where $\int_\omega \left[  \int_{\Omega(\chi_n,n)}{\cal D}z \right]  {\rm d}\chi_n$ simplifies to $ \int_\Omega {\cal D}z$ in Eq.~\ref{sumdirac}, with integration running over the space of either autonomous or non-autonomous paths ($\Omega=\Omega_{\rm a}$ or $\Omega_{\rm na}$). After permuting summation and integration in Eq.~\ref{sumdirac}, we introduced  in Eq.~\ref{sumdirac2} the posterior likelihood ${\rm P_{sel}}(\chi_n,n|z)$ by resorting to Bayes relation 
\begin{eqnarray}
{\rm P_{sel}}(\chi_n,n|z) =  \frac{{\rm P_{cond}} (z|\chi_n,n) {\rm P}^\varphi(\chi_n,n)}{{\rm P}^\varphi_{\rm M}(z)} \label{bayes}. 
\end{eqnarray}
The posterior probabilities can be evaluated and simulated like the conditional probabilities. Indeed, plugging the various reverse-to-forward probability ratios given by  
\begin{eqnarray}
\frac{{\rm P_{cond}} (z|\chi_0,0) {\rm P}^\varphi(\chi_0,0)}{{\rm P_{cond}} (z|\chi_n,n) {\rm P}^\varphi(\chi_n,n)} = \exp\left[\varphi(\chi_0)-\varphi(\chi_n) - \beta W_n \right] \label{pselect} \label{posterior}
\end{eqnarray}
into~\eqref{bayes}, yields the evaluable ratio  
\begin{eqnarray}
{\rm P_{sel}}(\chi_n,n|z) =  \frac{\exp\left[\varphi_0-\varphi_n - \beta W_n \right]}{\sum_{k=0}^N\exp\left[\varphi_0-\varphi_k - \beta W_k \right]} \label{pseleval}
\end{eqnarray}
where $W_n$ is given in~\eqref{effectivework2} and $\varphi_k$ stands for $\varphi(\chi_k)$ for simplifying. 

The aforementioned feature of the posterior and conditional probabilities makes it possible to generate a path distribution according to the marginal probability ${\rm P}^\varphi_{\rm M}(z)$. To explain how this can be done, let us consider a Monte Carlo move from $\chi_n$ to $\chi'_n$ and whose associated transition probability ${\rm P_{trans}} (\chi'_n|\chi_n)$ obeys a detailed balance with respect to ${\rm P}^\varphi$ given by~\eqref{toto} 
\begin{eqnarray} 
{\rm P_{trans}} (\chi'_n|\chi_n)  {\rm P}^\varphi(\chi_n,n) = {\rm P_{trans}} (\chi_n|\chi'_n)  {\rm P}^\varphi(\chi'_n,n) \label{toto} .  
\end{eqnarray} 
Then, considering a path $z'$ containing $(\chi'_n,n)$ and plugging Bayes relation~\eqref{bayes} into~\eqref{toto} for paths $z$ and $z'$ implies the detailed balance condition 
\begin{eqnarray} 
{\rm P_{cond}}(z'|\chi_n',n){\rm P_{trans}} (\chi'_n|\chi_n) {\rm P_{sel}}(\chi_n,n|z) {\rm P}^\varphi_{\rm M}(z) = {\rm P_{cond}}(z|\chi_n,n){\rm P_{trans}} (\chi_n|\chi_n') {\rm P_{sel}}(\chi'_n,n|z') {\rm P}^\varphi_{\rm M}(z') \label{webDB} 
\end{eqnarray}
in which ${\rm P_{cond}}(z'|\chi_n',n){\rm P_{trans}} (\chi'_n|\chi_n)  {\rm P_{sel}}(\chi_n,n|z)$ and ${\rm P_{cond}}(z|\chi_n,n){\rm P_{trans}} (\chi_n|\chi_n')  {\rm P_{sel}}(\chi'_n,n|z')$ have to be read as the probabilities to transit from path $z$ to $z'$ and from $z'$ to $z$, respectively. The path distribution generated by any sampler satisfying the detailed balance condition~\ref{webDB} is ensured to converge toward the probability distribution ${\rm P}^\varphi_{\rm M}$. 

Still, the canonical average~\eqref{canonical} of quantity $A(\br)$ must be extended in order to be evaluable from a sample of paths distributed according to ${\rm P}^\varphi_{\rm M}$. To achieve this task, we first write the canonical average with respect to the biased probability measure. From the relation 
\begin{equation} 
\frac{\rho (\chi)}{\rho^\varphi(\chi)} = \exp\left[\beta\left(\cF-\cF^\varphi 
\right) + \varphi(\chi)\right], \nonumber
\end{equation}
we obtain  
\begin{eqnarray} 
\left\langle A \right\rangle & = &\bigg[ \int_\omega A(\br) 
\frac{\rho(\chi)}{\rho^\varphi(\chi)}\rho^\varphi(\chi) {\rm d}\chi \bigg] {\bigg/} 
\left[ \int_\omega \frac{\rho(\chi)}{\rho^\varphi(\chi)}\rho^\varphi(\chi) {\rm d}\chi \right]  \nonumber \\ 
&= & \left[ \int_\omega A(\br) {\rm e}^{\varphi(\chi)} \rho^\varphi(\chi) {\rm 
d}\chi \right] \bigg/\left[ \int_\omega {\rm e}^{\varphi(\chi)} \rho^\varphi(\chi) 
{\rm d}\chi \right] \label{biasaverage}. 
\end{eqnarray}
Then, the path-integral expression of $\rho^\varphi$ (Eq.~\ref{sumdirac2}) is inserted into the biased average (Eq.~\ref{biasaverage}) and the Dirac functions are evaluated when integrating $\chi$ over $\omega$, which yields 
\begin{eqnarray}
\hspace{-2mm}\left\langle A\right\rangle
\hspace{-2mm}&= &\hspace{-2mm} \Bigg[ \int_\Omega \bigg[ \left.\sum_{n=0}^N \right. A(\br_n) {\rm e}^{\varphi(\chi_n)} {\rm P}_{\rm sel}(\chi_{n},n|z) \bigg] {\rm P}^\varphi_{\rm M}(z) {\cal D}z \Bigg] \Bigg/ \Bigg[ \int_\Omega \bigg[ \left. \sum_{n=0}^N \right.{\rm e}^{\varphi(\chi_n)} {\rm P}_{\rm sel} (\chi_n,n|z) \bigg] {\rm P}^\varphi_{\rm M}(z) {\cal D}z \Bigg] 
\label{webaverage}. 
\end{eqnarray} 
We are now in the position of evaluating the canonical ensemble average~\eqref{webaverage} by following the traditional recipe~\cite{frenkel:2002} : i) we construct a Markov Chain distributed according to the probability density ${\rm P}^\varphi_{\rm M}$ and ii) resort to the appropriate estimator to correct for the bias introduced by the importance function ${\rm P}^\varphi_{\rm M}$. 
\begin{description}
\item (i) Repeat $M$ times the following steps : 
\begin{description}
\item (a) move to the shooting index $n$ and state $\chi_n$ of current path $z$ which have been selected both with posterior likelihood ${\rm P_{sel}}(\chi_n,n|z)$ in the previous steps (d) and (e); 
\item (b) generate the new shooting state $\chi'_n$ from $\chi_n$ from probability ${\rm P_{trans}} (\chi'_n|\chi_n)$;
\item (c) set $C_0$ to $1$, initialize the (provisional) next shooting index $n_{\rm prov}$ to $n$ and store $\chi'_n$ into  $\chi_{\rm prov}$;
\item (d) shoot a new trajectory from state $\chi'_n$ and index $n$ (perform $N-n$ updates from $(\chi'_n,n)$ followed by $n$ downdates from $(\chi'_n,n)$ again); concomitantly, at each Langevin iteration $k$, compute $C_{h}=C_{h-1}+\exp\left[\beta (W'_n-W'_{\alpha(h)}) -\varphi'_{\alpha(h)} \right]$ where index $h=\alpha^{-1}(k)$ runs from $1$ to $N$ after re indexing using $\alpha$; with probability $1-C_{h-1}/C_h$, change $n_{\rm prov}$ to $\alpha(h)=k$ and store $\chi'_{\alpha(h)}$ into $\chi_{\rm prov}$, otherwise leave $n_{\rm prov}$ and $\chi_{\rm prov}$ unchanged; 
\item (e) set the next shooting index $n'$ to $n_{\rm prov}$ and store $\chi_{\rm prov}$ into $\chi'_{n'}$ to denote the selected state of the new completed path $z'$; 
\item (f) go to (a) until the chain has been completed.  
\end{description}
\item (ii) Evaluate the estimator
 \begin{eqnarray}
\hat{A}_M & = & \frac{ 
\sum_{m=1}^M \frac{\sum_{n=0}^N A_{n|m} \exp \left( -\beta {W}_{n|m}\right) } {\sum_{n=0}^N \exp \left( -\varphi_{n|m}-\beta {W}_{n|m}\right)}} 
{\sum_{m=1}^M\frac{\sum_{n=0}^N \exp\left(-\beta {W}_{n|m}\right) }{\sum_{n=0}^N \exp\left(-\varphi_{n|m}-\beta{W}_{n|m}\right) }} \label{heataverage} . 
\end{eqnarray}
where $\left\lbrace z_1,...,z_m,...,z_M \right\rbrace$ denotes the paths of the Markov chain constructed using the sampler and the simplified notations $A_{n|m}$, $\varphi_{n|m}$ stand for $A(\br_{n})$, $\varphi(\br_{n})$ of path $z_m$.  $W_{n|m}$ represents the work done upon the extended system along the trajectory $z_m$ between $\chi_{0}$ and $\chi_{n}$ (i.e. from times $t_0$ to $t_n$) via the mechanical coupling. 
\end{description}
The reindexing function is 
\begin{eqnarray}
\alpha(h) = 
\begin{cases} 
h+n   & \text{  if  } h \leq N-n, \\ 
N-h   & \text{  if  } h > N-n, 
\end{cases}
\hspace{2cm}
\alpha^{-1}(k) =
\begin{cases} 
k-n   & \text{  if  } k \geq n, \\ 
N-k   & \text{  if  } k < n. 
\end{cases} \nonumber
 \end{eqnarray}
Some details of the algorithm above such as the move from $\chi_n$ to $\chi'_n$ in step (i)-(b) depends on the specific implementation: we choose ${\rm P_{trans}} (\chi'_n|\chi_n)=\delta(\chi'_n-\chi_n)$ in Sections~\ref{application1D} implying that $\chi_n$ is left unchanged; in Section~\ref{application2D}, $\chi'_n$ is constructed from $\chi_n$ by drawing new momenta $\bp'_n$ from the Maxwell-Boltzmann distribution. 
Besides, the estimator~\eqref{heataverage} is obtained by plugging~\eqref{pseleval} into the usual Metropolis estimator  
\begin{eqnarray}
\hat{A}_M & = & \frac{\sum_{m=1}^M \sum_{n=0}^N A_{n|m} {\rm e}^{\varphi_{n|m}} {\rm P}_{\rm sel}(\chi_{n|m},n|m|z_m) }{\sum_{m=1}^M \sum_{n=0}^N {\rm e}^{\varphi_{n|m}} {\rm P}_{\rm sel}(\chi_{n|m},n|m|z_m)} 
\label{retrieving} 
\end{eqnarray}
related to ensemble average~\eqref{webaverage}. 
Additionally, the shooting move~\cite{dellago:2002} of step (i-d) generates the new path $z'$ with probability ${\rm P}_{\rm cond}(z'|\chi'_n,n)$ given in Eq.~\ref{shooting} by construction. The next shooting state $\chi'_{n'}$ and next shooting index $n'$  obtained from  $\chi_{\rm prov}$ and $n_{\rm prov}$ are eventually selected with the compound probability ($0 \leq h' \leq N $)
\begin{eqnarray}
\left\{ \prod_{h=0}^{\max(0,h'-1)}\left[\frac{C_{h-1}}{C_h}+(1-\frac{C_{h-1}}{C_h}) \right] \right\} \left[1-\frac{C_{h'-1}}{C_{h'}} \right] \left\{ \prod_{h=h'+1}^{N} \frac{C_{h-1}}{C_h}  \right\} =\frac{C_{h'}-C_{h'-1}}{C_N}  \nonumber
\end{eqnarray}
where $C_{-1}=0$ and $h'=\alpha^{-1}(n')$. This compound probability is equal to ($n'=\alpha(h')$)  
\begin{eqnarray}
\frac{C_{h'}-C_{h'-1}}{C_N} = \frac{\exp\left[-\varphi'_{n'} - \beta W'_{n'} \right]}{\sum_{k=0}^N\exp\left[-\varphi'_k - \beta W'_k \right]} = {\rm P_{sel}}(\chi'_{n'},n'|z').  \nonumber
\end{eqnarray}
As a result, the algorithm satisfies the detailed balance equation~\eqref{webDB}. Note that the decomposition of the selecting procedure in (c-e) of (i) avoids storing all the configurations when a new path is constructed. 

\subsection{Conditional expectation viewpoint~\label{alternative}} 

In the residence weight algorithm, the shooting index $n'$ related to path $z'$ subsequent to $z$ is constructed concomitantly with $z'$ as outlined by steps (i)-(c) to (i)-(e) of subsection \ref{rwa}. As a result, the algorithm  satisfies the following detailed balance condition
\begin{eqnarray} 
{\rm P_{sel}}(\chi'_{n'},n'|z'){\rm P}_{\rm cond}(z'|\chi'_{n},n){\rm P}^\varphi(\chi'_{n},n) & = & 
{\rm P_{sel}}(\chi'_n,n|z') {\rm P}_{\rm cond}(z'|\chi'_{n'},n'){\rm P}^\varphi(\chi'_{n'},n'), \label{waste-recycling}
\end{eqnarray} 
for moves between $(\chi'_{n},n)$ and $(\chi'_{n'},n')$, and also obeys detailed balance equation~\eqref{toto} for moves between $(\chi_{n},n)$ and $(\chi'_{n},n)$. The algorithm thus leaves invariant the prior probability density ${\rm P}^\varphi$ : the Markov chain of states $(\chi_n,n)$ constructed with the RW algorithm is thus distributed according to the probability density ${\rm P}^\varphi$. 

It follows from this theoretical description that the residence weight algorithm is a particular implementation of the waste recycling algorithm introduced by Frenkel~\cite{frenkel:2004,athenes:2007}. 
Indeed, assuming there is no biasing potential in Eq.~\ref{retrieving}, one can subsequently normalize the selection probabilities to one ($M$ times) and write 
\begin{eqnarray}
\hat{A}_{M} = \frac{1}{M}\sum_{m=1}^M \sum_{n=0}^N A_{n|m} {\rm P_{sel}(\chi_{n|m},n|m|z_m)}. \label{waste-recycling}
\end{eqnarray}
which corresponds to the waste-recycling estimator given in Eq. 2.2 of Ref.~\cite{frenkel:2004}. Our symmetric selection procedure ${\rm P}_{\rm sel}$ corresponds to a Barker acceptance rule (\cite[Eq. 2.1]{frenkel:2004}) that considers states linked by the trajectories, while the wasted information included in the rejected Monte Carlo moves is recycled in the estimator of Eq.~\ref{waste-recycling}.  

Interestingly, Delmas and Jourdain~\cite{delmas:2007} showed that the estimator can be interpreted as the conditional expectation of $A$ with respect to ${\rm P}_{\rm sel}$ and that it behaves normally asymptotically. Additionally, these authors proved that the statistical variance of the estimator is smaller than the one of the Metropolis-Hasting estimator when the acceptance rule is symmetric~\cite{delmas:2007}, as in the present situation. 
The first property implies that the conditional estimator is unbiased whatever the sample size $M$, at variance with maximum-likelihood estimators that are only unbiased in the limit of large samples~\cite{chodera:2008}. The last property involving the variance reduction justifies the present strategy : when possible, one should systematically include the information contained in the states of the paths in statistical path-averages. 

We now turn to the applications of our method. In Section~\ref{application1D}, we implement non-autonomous steering along a one-dimensional order parameter with no biasing potential ($\varphi=0$) and will resort to Eq.~\ref{waste-recycling} to reconstruct the free energy profile. Then, in Section~\ref{application2D}, the method will be tested on a more difficult benchmark model involving the direct reconstruction of a two-dimensional free-energy landscape. This task will be achieved by resorting to autonomous steering with respect to two additional collective variables. In addition, the biasing potential $\varphi$ will be constructed iteratively. 

\section{1D free-energy reconstruction using non-autonomous steering} \label{application1D}

The one-dimensional reconstruction problem involves calculating the migration free-energy  of the vacancy in the $\alpha$-Fe system. 
Atomic interactions of the model system are described by the (embedded atom method) potentials developed by Ackland and coworkers~\cite{ackland:2004} and are computed using the minimum image convention. The crystal structure is body-centered cubic and the initial unrelaxed cell contains 1023 atoms displayed on 1024 lattice sites : the vacant site (vacancy) is at a corner of the cell. Let $\br_1=(q_1,q_2,q_3)$ denote a nearest neigbour atom of the vacancy along a [111] direction and define the one-dimensional collective variable ${\rm dist }(\br_1)$ as the distance between $\br_1$ and the system's center of mass $\mathbf{R}_m$. A single additional variable $q_{3I+1}$ is associated to ${\rm dist }(\br_1)$ via the potential energy 
$\frac{\kappa}{2}(q_{3I+1}-{\rm dist }(\br_1))^2$. The time-step is $\Delta t=4 \cdot 10^{-15}s$ and the friction of the Langevin dynamics is $\gamma_i = 2.5 \cdot 10^{12} s^{-1}$ whatever $i$. 

Furthermore, protective spheres have been added upon the 7 nearest atoms of atom $\br_1$ and upon the 7 nearest atoms of the vacancy ($\br_1$ being obviously unprotected). Each sphere is centred on the corresponding site of the underlying rigid lattice and is of radius $a/2$ (where $a=2.4728 \cdot 10^{-10} m$ is the nearest neigbour distance). Displacements moving a neigbouring atom out of its protective sphere are discarded, which can be done because the dynamics have been metropolized (see Appendix~\ref{Langevin}). The exit frequency of neighbouring atoms remains negligible even at high temperature. This procedure prevents spontaneous vacancy atom exchanges that may occur at temperatures above $540K$ by the nonequilibrium steering without altering the statistics. 

The reaction coordinate $\xi(\br)$ is the projection of vector $\br_1-\mathbf{R}_m$ along [111] direction. Measuring the quantity of interest, ${\rm P}(\xi)$, via a histogram amounts to monitoring the occupation probability of atom $\br_1$ along [111] direction. 
We implement non-autonomous steering with its additional variable ($\xi^{\rm add}=q_{3I+1}$) evolving at constant velocity according to 
\begin{eqnarray}
q_{3I+1}(t_n) = q_{3I+1}(0) + \frac{t_n}{\tau} \left[  q_{3I+1}(t_N) - q_{3I+1}(0)\right].  
\end{eqnarray}
The values $q_{3I+1}(0)=-\frac{a}{10}$ and $q_{3I+1}(\tau)=\frac{11a}{10}$ in the steering schedule have been chosen such that the atom $\br_1$ performs a single jump into the vacant site. The phase space is thus restricted to 2 possible vacancy sites. 
Time is given by $t_n=n \Delta t$. 
As noticed in related studies~\cite{bruneval:2007,adjanor:2008} implementing Jarzynski's work identity, we found it advantageous to use few long nonequilibrium trajectories rather than many short ones. Hence, for each temperature, we have generated $M=100$ hybrid trajectories with $N=10^{5}$ time-steps (implying a total of $10^7$ force evaluations per simulation). From the 100 trajectories, we calculated the histograms $\widehat{\rm P}_{100}(\xi)$ with $\xi$ spanning the interval $\left[ -\frac{a}{10},\frac{11a}{10} \right]$ in 121 bins. Temperature ranges from 20K to 1000K. 

\begin{figure}
\vspace{-4cm}
\scalebox{2}{\includegraphics{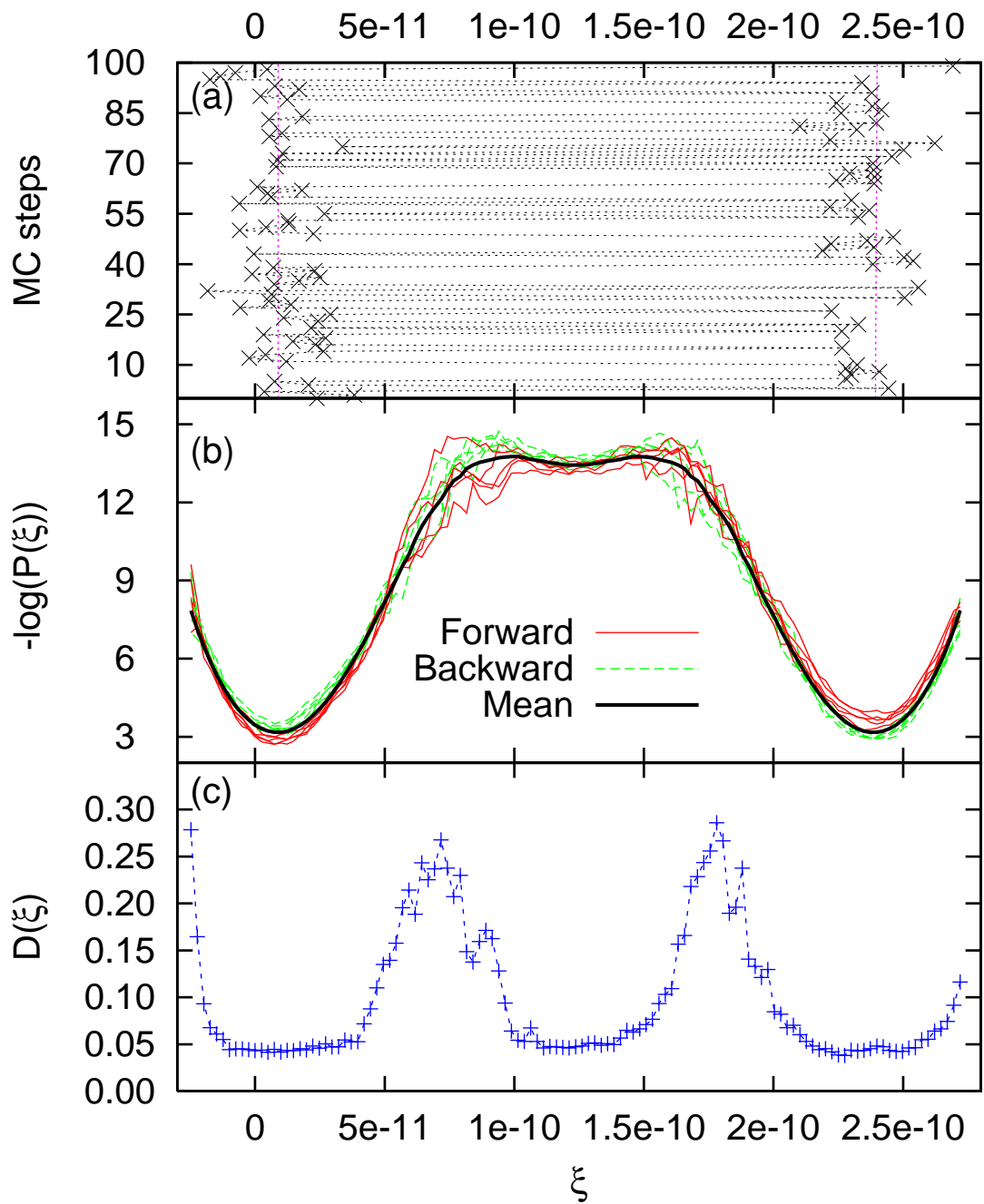}}
\caption{Various simulation outputs expressed as a function of the reaction coordinate $\xi$  in meters. Panel (a) displays the trajectory number versus the $\xi$ value of the selected states ($\times$); panel (b) represents the colog of ten $\widehat{P}_1(\xi)$ estimates measured from single paths generated either forward (red thin curve) or backward (green dashed curve); Panel (c) represents the divergence of the estimates. Since the biasing potential $\varphi$ is zero, the selected reaction coordinates are distributed according to the equilibrium distribution ${\rm P}(\xi)$ itself. This one exhibits two maxima located by the position of the vertical dashed lines. 
The dashed segments that join the $\times$ symbols represent pairs of successively selected states. The fraction of dashed segments 
crossing the symmetric free energy barrier is about 40\% at the considered temperature.}
\label{various_estimates}
\end{figure}

Figure~\ref{various_estimates} displays various outputs of the simulation carried out at $T= 540 K$. Panel (a) displays the reaction coordinates $\xi$ of the states successively selected by ${\rm P}_{\rm sel}$ (Eq.~\ref{pselect}). The high crossing probability discussed in the figure caption is related to the small value of the work performed on the system once the system has transited over the barrier at the present low-velocity steering. 
To illustrate the unavoidable lag effect caused by steering, let us consider the cologarithm of a $\widehat{P}_1(\xi)$ estimate (i.e. of an estimate obtained from a single trajectory). The variation of $-\beta^{-1} \ln \widehat{P}_1(\xi)$ with respect to $\xi$ is related to the work done along the trajectory. We have represented in panel (b) the cologarithm of several $\widehat{P}_1(\xi)$ estimates for forward or backward trajectories. The asymmetry of the colog-probability profiles is controlled by the steering direction and the amount of dissipation. 
The asymmetry is removed because the estimator combines trajectories generated forward and backward along $\xi$, i.e. starting from both free energy minima. The cologarithm of the $\widehat{\rm P}_{100}(\xi)$ histogram, represented in panel (b) by the thick symmetric curve, illustrates this feature. A similar compensation has been observed previously~\cite{minh:2008} with a bidirectional variant of the Hummer-Szabo method. In panel (c), we have plotted  the divergence defined by  
\begin{eqnarray}
D(\xi) = \ln P(\xi)-\overline{\ln \widehat{P}_1(\xi)}
\end{eqnarray}
where $P(\xi)$ is estimated here using $\widehat{P}_{100}(\xi)$ and the overbar denotes averaging the 100 available $\ln \widehat{P}_1(\xi)$ values. 
Mathematically, the divergence $D(\xi)$ is a relative entropy betwen two distributions~\cite{jarzynski:2006}. Thermodynamically, it is an excess entropy that is stored into the system and that would be irreversibly dissipated toward the thermostat if the system was allowed to relax back to equilibrium at constant $\xi$. This quantity gives information on the convergence of exponential averages~\cite{athenes:2004,jarzynski:2006}. The smaller the divergence is and the more accurate the estimation is. We observe from Fig.\ref{various_estimates} that accuracy is smaller at the edges of the barrier, where the gradient of the steering potential is larger. Besides, $D(\xi)$ decreases again to a minimum around $\xi=1.6 \cdot 10^{-10}m$ that corresponds to the intermediate free energy minimum of panel (b). This trend suggests that the excess energy transiently stored in the steering potential is not entirely dissipated but is released to the extended system when $\xi$ reaches the intermediate energy minimum. 

\begin{figure}
\vspace{-3cm}
\hspace{1cm}\scalebox{1}{\includegraphics{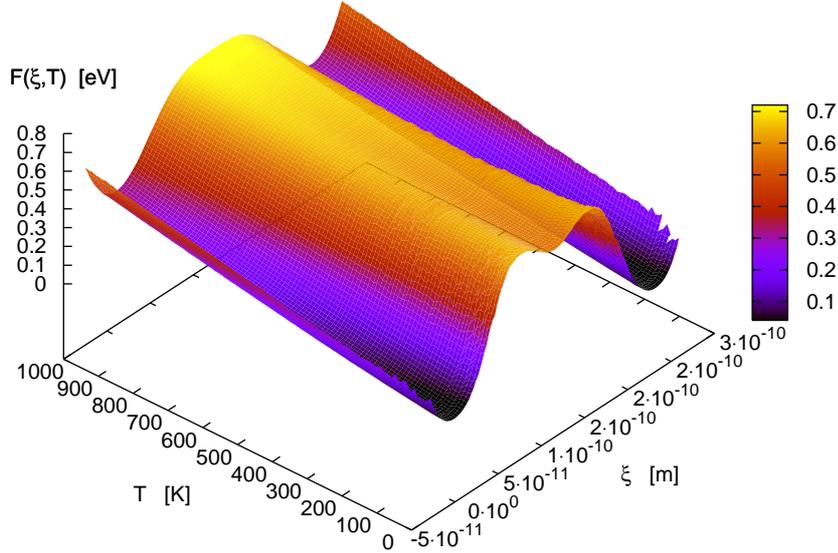}}
\caption{
Free energy $F(\xi,T)$ as a function of $\xi$ and $T$, estimated from $-kT \ln \hat{P}_{100}$ for the corresponding temperature. }
\label{free_energy_landscape}
\end{figure}
\begin{figure}
\vspace{-1cm}
\hspace{1cm}\scalebox{1}{\includegraphics{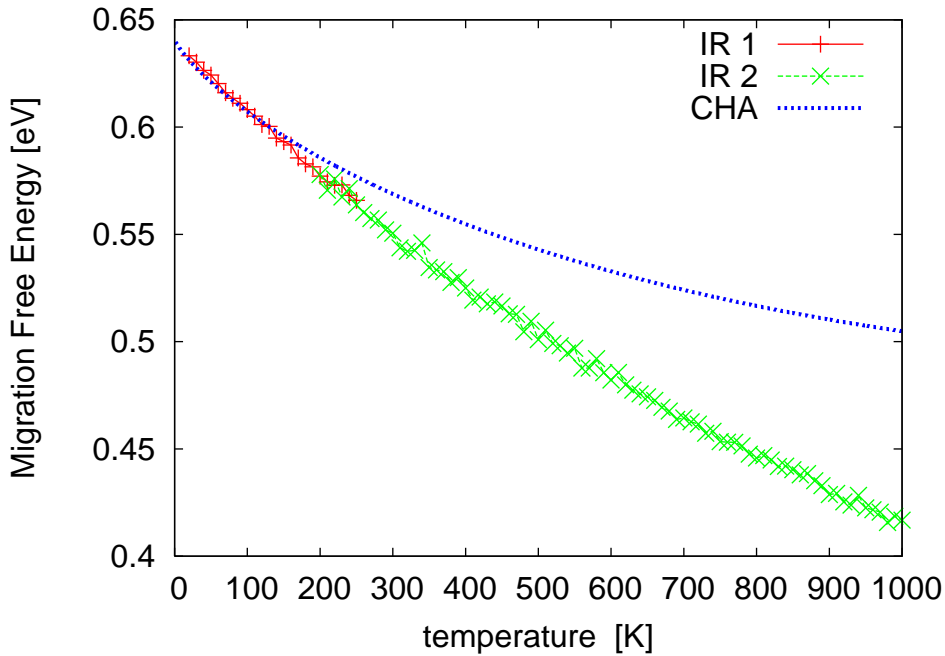}}
\caption{Migration free energy as a function of temperature. IR1 and IR2 refer to two values of $\kappa$ (see text). The dotted line are the results of classical harmonic approximation. }
\label{comparison_vacancy}
\end{figure}

The method was found to yield reproducible $F(\xi,T)$-estimates down to the temperature of $20K$. Two series of simulations were carried out. From $20K$ to $250K$, we used $\kappa=10\cdot 10^4 J\cdot m^{-2}$ (IR1) and from $200K$ to $1000K$ we used $\kappa=5\cdot 10^4 J\cdot m^{-2}$ (IR2). 
Results are represented by the free energy landscape of Fig.~\ref{free_energy_landscape}. We observe that the intermediate free-energy minimum is more pronounced at the lower temperatures and completely disappears at temperatures higher than $700K$. The migration free energies are deduced from the relative barrier heights along $\xi$-axis of  Fig.~\ref{free_energy_landscape}. They are plotted as a function of temperature in Fig.~\ref{comparison_vacancy} together with the prediction of classical harmonic  approximation (CHA). CHA calculations have been performed 
using the procedure described in  Ref.~\cite{athenes:2006, marinica:2007} and considering one of the two symmetric energy minima and saddle configurations. As expected, Monte Carlo simulations and CHA calculations agree  at low temperatures ($T<200K$) where anharmonic effects are negligible, confirming the  exactness of our simulation method. At temperatures higher than $200K$, we  observe a substantial deviation between simulation and CHA, attesting to strong an anharmonicity.  Note that the extent of anharmonicity is in quantitative agreement with the one previously reported in the literature~\cite{marchese:1986} for the vacancy migration free energy. 

In our first application, a single collective variable was used and simulations were performed successively with varying the temperatures so as to complete the landscape. In the second application, we show how to achieve two-dimensional reconstruction directly by resorting to autonomous steering with two additional variables.  

\section{2D free-energy reconstruction using autonomous steering } \label{application2D}

\begin{figure}
\vspace{-0.5cm}
\hspace{2cm}\scalebox{0.3}{\includegraphics{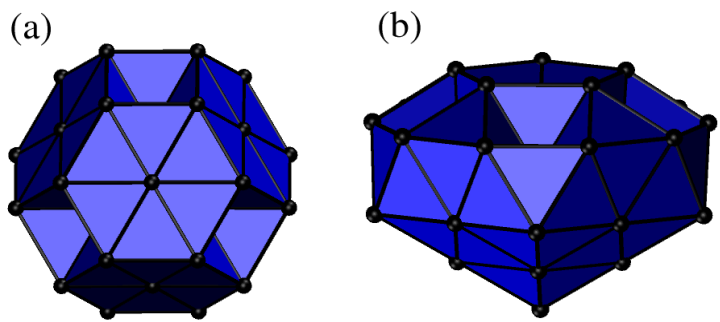}}
\caption{The two lowest energy structures of the 38-atom cluster: 
(a) truncated octahedron with energy $E_0=-173.9284$ and 
(b) incomplete icosahedron, $E_1 = -173.2524$.}
\label{lj38}
\end{figure}

We consider the 38-atom Lennard-Jones cluster. 
LJ$_{38}$ is computationally troublesome to study because its potential energy landscape has two main funnels~\cite{Doye:1999,CalvoNFD00,Bogdan:2006}, whose respective lowest energy structures are the icosahedron and the octahedron displayed in Fig.~\ref{lj38}. It undergoes a two-stage phase change with increasing temperature starting from the octahedral structure. 
A solid-solid transition temperature between the octahedral funnel and the icosahedral funnel occurs near $T_{ss}\approx 0.12 \epsilon/k_B$, melting follows near $T_{ls} \approx 0.17 \epsilon/k_B$. LJ reduced units of length, energy and mass ($\sigma=1$, $\epsilon/k_B=1$, $m=1$) will be used in the following. 

\begin{figure}
\vspace{-2cm}
\hspace{1.5cm}\scalebox{1}{\includegraphics{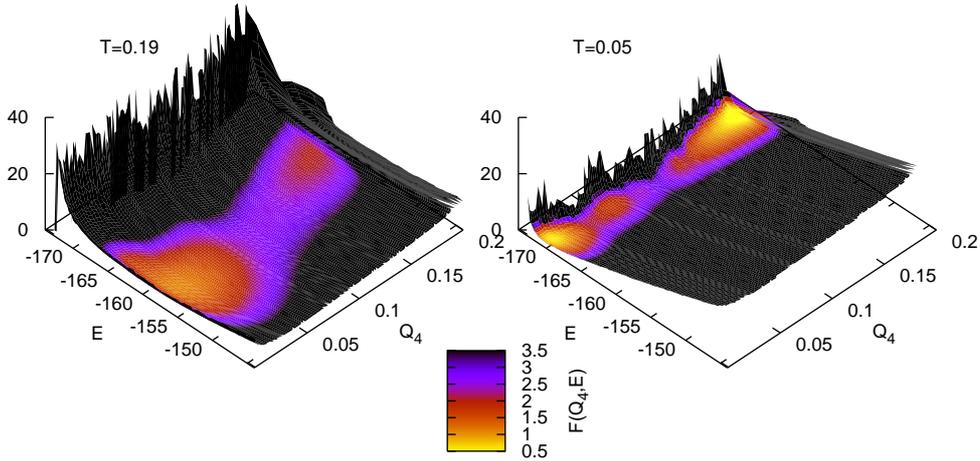}}
\vspace{-2cm} \caption{Free energy $F(Q_4,E)$ as a function of $Q_4$ and $E$. Left panel is the actual measurement at $T=0.19$, while the right panel represents the free energy reconstruction for temperature $T=0.05$ as obtained by Legendre transform. }
\label{contour}
\end{figure}

As first collective variable, we use the bond-orientational order parameter $Q_4$ of Steinhardt et al. ~\cite{Steinhardt:1983} that is able to distinguish between the cubic structures favoured at low temperatures and the icosahedral isomers above $T_{ss}$. The second collective variable is the potential energy $E (\br)$. 
The associated additional variables, $q_{3I+1}$ and $q_{3I+2}$, act upon the particles via  harmonic potentials whose stiffnesses are $\kappa_1=10^{4}$ and $\kappa_2=2$, respectively. Their respective masses are $m_{3I+1}=6400$ and $m_{3I+2}=0.8$. The respective coupling parameters are  $\mu_{3I+1}=0.9$ and $\mu_{3I+2}=0.995$, and the frictions are $\gamma_{i}=5 \cdot 10^{3}$ ($i \le 3I$) and $\gamma_{j}=5 \cdot 10^{-3}$ ($j>3I$). They have been chosen using the simple recipe that follows ($j > 3I$) : (i) the $\mu_j$'s are tuned to enable the additional variables $q_j$ to oscillate with an amplitude large enough in the direction of the corresponding order parameter; (ii) the masses $m_j$ are then tuned to set the velocity slow enough (but not too slow) and (iii) the coupling parameter $\gamma_j$ is chosen small enough to ensure a smooth and regular evolution of the $q_j$'s. Procedures (ii) and (iii) prevent the dynamics from producing entropy, i.e. from dissipating the work done on the system into heat when the $q_j$'s evolve too fast. The values given above were found satisfactory and are certainly sub-optimal. Finding the optimal computational set-up is a non trivial task. 

A series of iterative simulations have been carried out at the temperature $T=0.19$, 
using the procedure introduced by Coluzza and Frenkel~\cite{coluzza:2005} in a similar context. Let ${\rm P}^\ell(Q_4,E)$ denote the histogram constructed by the $\ell$-th simulation. The  biasing potential $\varphi^{\ell+1}$ of the next simulation is then constructed using the  iterative procedure 
\begin{eqnarray}
\varphi^{\ell+1}=\phi^{\ell+1} \circ \bs \hspace{1cm} \phi^{\ell+1}(Q_4,E) = -\ln(\widehat{\rm P}^\ell(Q_4,E)+p^\ell_{\rm min}). 
\end{eqnarray}
The $p^\ell_{\rm min}$ parameter determines the maximum value of the biasing potential and  thus avoids possible singularities arising from unexplored histogram bins. As the successive  simulations explore larger portions of the phase space more and more accurately, the control parameter  is decreased iteratively using the relation $p^\ell_{\rm min}=10^{-9-4\ell}$. Each simulation generates approximately $M=10^5$ trajectories of $N=2.5\cdot 10^5$ time-steps.

\begin{figure}
\vspace{-2cm}
\hspace{3.5cm}\scalebox{1.3}{\includegraphics{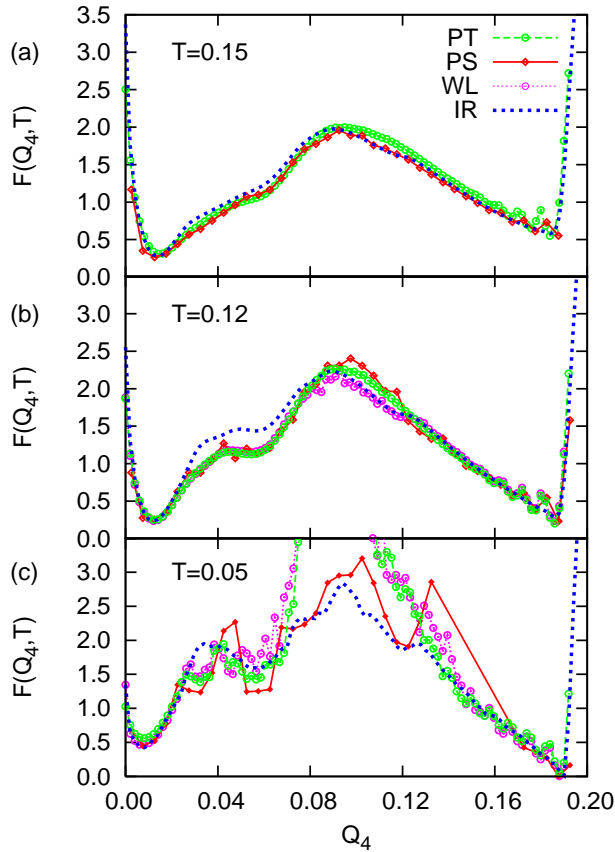}}
\caption{Free energy profiles of LJ$_{38}$ as a function of $Q_4$ obtained with information-retrieval (IR), path-sampling (PS), parallel tempering (PT), Wang-Landau (WL). (a) $T=0.15$ ; (b) $T=0.12$ ; (c) $T=0.05$.}
\label{comparison}
\end{figure}

Panel (a) of figure~\ref{contour} represents the final $F_{T_0}(E,Q_4)$ contour plot obtained at the temperature of the simulation $T_0=0.19$. 
The color scale is such that the improbable regions of the landscape are displayed in black. 
The free energy at any temperature $T_1$ is related to the microcanonical density of states $g(Q_4,E)$ via the relation~\cite{Wales:2003} 
\begin{eqnarray}
\frac{F_{T_1}(Q_4,E)}{k_B T_1} & = & \frac{E}{k_BT_1}- \ln g(Q_4,E) + \ln \int g(Q_4,E) \exp \left[-\frac{E}{k_BT_1} \right] dE. 
\end{eqnarray}
Since this relation is also valid for the reference temperature $T_0$, it gives access to the  difference of Landau free energies between any target temperature $T_1$ and the reference  temperature $T_0$ of the simulation 
\begin{eqnarray}
\frac{F_{T_1}(Q_4,E)}{k_B T_1} & = & \frac{E}{k_BT_1} -\frac{E-F_{T_0}(Q_4,E)}{k_B T_0} + \ln \int \exp\left[\frac{E-F_{T_0}(Q_4,E)}{k_B T_0} -\frac{E}{k_BT_1} \right] dE.  \label{intoverE}
\end{eqnarray}
The integral over energy in~\eqref{intoverE} acts as a normalizing factor and is the partition-function ratio  involving the reference ($T_0$) and target ($T_1$) systems. The free energy landscape at  temperature $T_1=0.05$ is finally displayed in panel (b) of Fig.~\ref{contour} and  reveals the low energy structures previously reported in Ref.~\cite{athenes:2006}. To make a quantitative comparison between the present method [information retrieval with autonomous-steering, (IR)] and the three simulation methods used in Ref.~\cite{athenes:2006} [nonequilibrium path-sampling (PS), parallel-tempering (PT) and Wang-Landau (WL)], we plot in Fig.~\ref{comparison} all the estimated free energy profiles, $F(Q_4,T)$ as a function of $Q_4$, for temperatures $T=0.15$, $0.12$ and $0.05$. 

For the present method and the Wang-Landau method~\cite{athenes:2006}, we used the standard relation 
\begin{eqnarray}
F(Q_4,T) = -k_BT \ln \int \exp\left[\frac{-F_{T}(Q_4,E)} {k_B T} \right] dE
\end{eqnarray}
to obtain the free energy profile. 
In the reported path-sampling simulations, the temperature along the trajectories were slowly cooled down starting from $T=0.19$, which, using the present terminology, amounts to non-autonomous steering with respect to temperature. However, in contrast with the present study, the estimator that was implemented was based on Crook's nonequilibrium average and could only exploit the information from the time-slice of the corresponding temperature. 

We observe that, at the temperature $T=0.15$, the lowest that could be simulated correctly using standard umbrella sampling and histogram reweighting~\cite{Doye:1999}, our IR estimates for the free energies are in excellent agreement with the PS, PT and WL estimates~\cite{athenes:2006}. 
However, at lower temperatures, we observe a disagreement for low free-energy structures in the range $0.02 \leq Q_4 \leq 0.07$, in particular for the one appearing around $Q_4\approx 0.03$, compared to estimates obtained using the three other methods (PS, PT and WL). This limitation of the method is con commitant to the slow convergence of the two-dimensional biasing potential $\phi^\alpha$ in this region of phase space. We did not perform another iteration as the total number of force evaluations was already $10^{11}$, i.e. equal to the one in the corresponding PS simulations~\cite{athenes:2006}. 

Nevertheless, at temperature $T=0.05$ [Fig.\ref{comparison}, panel (c)], the estimates (IR) of $F(Q_4,T)$ revealed numerous low energy structures in the range $0.07 \leq Q_4 \leq 0.12$, which are explored with path-sampling, but missed with parallel tempering and Wang-Landau sampling. 
In particular, the low energy structure at $Q_4=0.12$ is correctly predicted. Interestingly, the bassins of attraction of these minima are larger with IR than with PS. This is due to the enhanced statistics made possible by the multi-state estimator that exploits information from all time-slices at any temperature, unlike the PS estimator. 

\begin{figure}
\vspace{-2cm}\hspace{-2cm}\scalebox{1.2}{\includegraphics{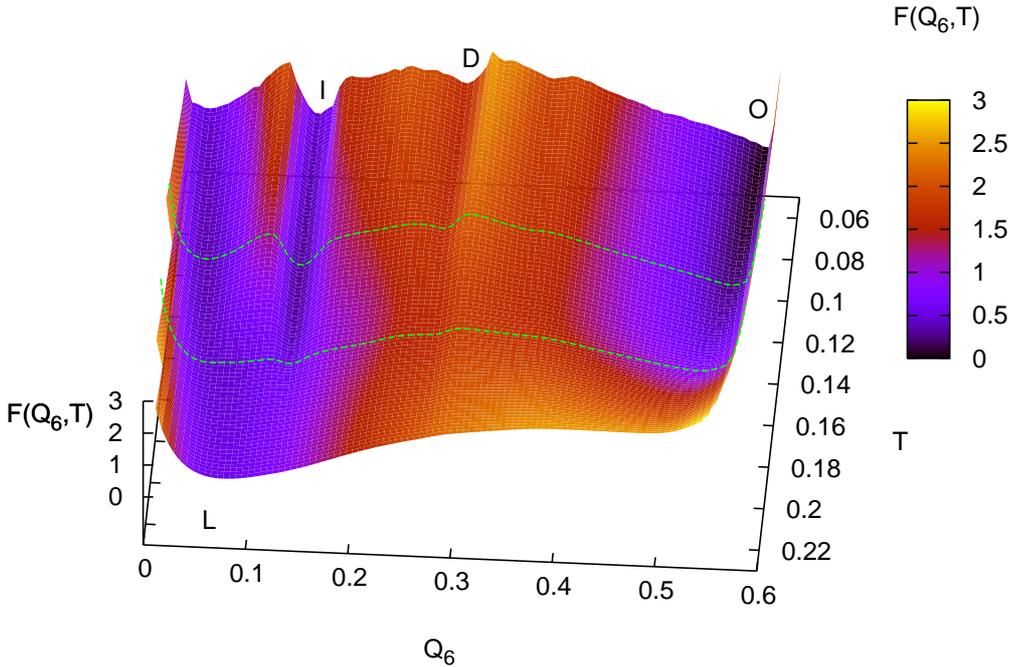}}
\caption{Free energy $F(Q_6,T)$ as a function of $Q_6$ and $T$. The dotted line at $T=017$ represents the phase transition between the liquid-like structure and the icosahedral structure (I). The one at $T=0.12$ shows the transition between the icosahedral structure and the octahedral structure (O). The label D represents defected structures near $Q6=0.25$ (or $Q_4=0.08$). }
\label{contour_q6}
\end{figure}
Finally, we plot  the two-dimensional free-energy landcape $F(Q_6,T)$ as a function of $Q_6$ and $T$ obtained from the simulation using a multi-state estimator in Fig.~\ref{contour_q6}. The $Q_6$ order parameter is used because it better distinguishes the liquid phase $L$ and the icosahedral phase $I$, as seen from this figure. Results are in qualitative agreement with previous simulation (PS) except for the defected structures $D$ around $Q_6 \approx 0.24$ whose free energy is overestimated. 

To conclude this test study on LJ$_{38}$ system, the reconstruction of a two-dimensional free-energy was achieved owing to the iterative construction of the two-dimensional biasing potential $\phi$ with autonomous steering. 
In term of numerical efficiency, the approach was found advantageous because its estimator retrieves the information contained in all the time-slices of the generated trajectories, unlike the nonequilibrium average previously implemented with non-autonomous steering~\cite{athenes:2006}. Further comparing the numerical efficiency of the methods from  Ref.~\cite{athenes:2006} and the present one is difficult, as estimators and steering schedules of distinct types were used. The non-autonomous steering schedule of Ref.~\cite{athenes:2006} should be tested with the more efficient multi-state estimator proposed in this study. In addition, given that the reported Wang-Landau sampling simulations achieved greater performance in the range $0.04 \leq Q_4 \leq 0.08$ but smaller performance in the range $0.08 \leq Q_4 \leq 0.14$, it may be worth implementing the multi-state estimator described in the present paper in combination with adaptive sampling methods such as  Wang-Landau sampling~\cite{wang:2001,calvo:2002} or metadynamics~\cite{laio:2002} so as to check whether the construction of the biasing potential would be facilitated. 

\section{Concluding remarks} \label{concluding}

In this article, we developed a unifying framework and a simple algorithm for retrieving the equilibrium information contained in all the time-slices from a sample of nonequilibrium trajectories. The algorithm, which shares several features with maximum-likelihood methods for nonequilibrium dynamics, is built upon Bayes theorem : a sampler generates a Markov chain of linked trajectories distributed according to a marginal probability, while an estimator operates over the Markov chain so as to infer the contribution to equilibrium of the sampled data using a likelihood function. Contrary to maximum-likelihood estimators, the proposed estimator does not involve post-processing, can possibly be implemented with dynamics based on either non-autonomous or autonomous scheduling and is unbiased but not optimal. 

Concerning the overall efficiency of the method, we observe that, in agreement with theoretical prediction~\cite{oberhofer:2008} and maximum-likelihood simulations~\cite{minh:2008}, the most accurate estimations of free energies are obtained when trajectories can be initiated from the various regions of interest. This computational requirement can be fulfilled by tuning an auxiliary biasing potential so as to flaten the prior probability distribution along the desired reaction coordinate. Using autonomous steering dynamics and resorting to a simple iterative procedure for constructing a two-dimensional biasing potential, the multi-state estimator could indeed reconstruct the free-energy landscape of the troublesome LJ$_{38}$ system quite accurately. 

The presented simulations clearly outlined the advantages of steering autonomously rather than non-autonomously : the former strategy can be implemented in complex systems that require more than one steering variable and enables one to reconstruct multi-dimensional free energy landscapes directly. This feature in fact extends the possibilities of the Hummer-Szabo methodology~\cite{hummer:2001}. 

Eventually, the multi-state estimator should be implemented with an adaptive sampler~\cite{lelievre:2007,barducci:2008} to check whether the combination of both techniques facilitates or not the construction of the biasing potential, compared to simulations resorting to one of the two techniques exclusively. This issue is to be considered in the wider perspective of waste-recycling~\cite{frenkel:2005,delmas:2007}, which similarly advocates to retrieve all the information generated during the simulations  within on-line statistical averages. A recent numerical investigation~\cite{athenes:2008c} shows that a significant reduction of the statistical variances can be achieved in replica-exchange simulations that implement a multi-state estimator and a multi-proposal sampler~\cite{esselink:1995} similar to the ones used in the present study. 

\section*{Acknowledgment}
We are much indebted to Eric Vanden-Eijnden for suggesting the use of autonomous steering~\cite{maragliano:2006} and for advising us. We warmly thank Gabriel Stoltz for relevant comments on early versions of the manuscript, Florent Calvo for providing us with the $Q_4$-gradient subroutine, Giovanni Ciccotti, John Chodera and David Minh for valuable advice. This work was mainly financed by Commissariat \`a L'Energie Atomique (DSOE program), partly carried out at Lawrence Livermore National Laboratory (under a VSP agreement) and performed using HPC resources from GENCI-CINES (Grant 2009-x2009096020). 

\appendix

\section{Residence algorithms and optimized path-sampling~\label{workbiased}}

In other implementations of the residence weight algorithm, the biasing function $\phi$ is a work quantity rather than an auxiliary potential as in the present study. The equivalent form of our detailed balance equation~\eqref{webDB} corresponds to the detailed balance condition~\cite[Eq.~61]{athenes:2007} which itself formalizes the weighted balance condition given in earlier works (refer to~\cite[Eq.~5]{athenes:2002a} and~\cite[Eq.~21]{athenes:2002b}). The weighted balance equation was induced by analogy with the residence time algorithm. The latter algorithm~\cite{lanore:1974,bortz:1975} and its extensions~\cite{athenes:1997,mason:2004,bulatov:2006} are used extensively in kinetic Monte Carlo simulations and also achieve importance sampling in ensembles of linked trajectories (more precisely, of kinetic pathways), owing to a similar selecting procedure. 
The selecting probability satisfies a detailed balance condition weighted by residence times (mean first passage times of exit). Eventually, the residence time algorithm also involves information retrieval~\cite{ceperley:1977,athenes:2004,adjanor:2005,athenes:2007}. The kinetic pathways that can possibly be constructed from the master equation and that are eventually discarded by the algorithm~\cite{wales:2009} do contribute to the residence times. These analogies are more obvious for the residence weight algorithms used in Ref.~\cite{athenes:2002a,athenes:2002b,athenes:2004,adjanor:2005,athenes:2007}, which unlike the present case, generate and select trajectories pertaining to ramified paths called webs~\cite{boulougouris:2005}. 

The relatively high numerical efficiency of the residence algorithms for estimating differences of free energies can be qualitatively explained by the study of Oberhofer and Dellago~\cite{oberhofer:2008}. Assuming that the biasing potential is an adjustable functional depending on the work ${\cal W}(z)$, these authors derived the optimal work dependent bias that leads to minimal statistical variance. The statistical variance was found to be minimal when the work-bias distribution contains typical forward and typical backward trajectories with similar weights, which is precisely an essential feature of the residence weight algorithms previously proposed. These algorithms generated paths using alternately the forward and reverse distribution in the extended ensemble of trajectories. 

In the present study, the biasing potential  $\varphi= \phi \circ \bl$ is state dependent. The appropriate potential ensuring equipartition of trajectories would be such that $\phi(\bl) \approx -\beta F(\bl)$, which amounts to artificially flattening the probability density along the order parameter $\bl$. Optimizing the biasing potential $\varphi$ requires knowledge of the quantities to compute, implying that the optimal bias can only be constructed iteratively, or adaptively, as outlined by Oberhofer and Dellago~\cite{oberhofer:2008}. 

\section{Metropolization of Langevin dynamics\label{Langevin}}
 
The unavoidable discretization errors are corrected in the ensemble average because the multi-state estimators uses the ratio of actual generating probabilities~\cite{athenes:2004}. Nevertheless, numerical efficiency depends on the choice of the time step $\Delta t$. Too small a time step  
decreases the sampling efficiency because states along the generated trajectories appear to be strongly correlated. Conversely, too large a time step produces numerical entropy (dissipated work~\cite{athenes:2004}) and one would observe that the selected states of the successive paths are also separated by small numbers of steps on average. 
Both situations results in increased statistical covariances in the path ensemble. Hence, in practice, the time step is tuned to achieve the best trade-off between decorrelation and entropy production. 

Unfortunately, for systems with many particles, discretization errors are important which imposes to choose a very small time step. 
This situation is encountered with our tabulated EAM potential of Iron. We have therefore metropolized the Langevin dynamics, i.e. we accept an iteration with probability~\cite{athenes:2006,bussi:2009} 
\[{\rm P}^{\rm acc}_{k}=\min(1,\exp\left[-\beta \left( \cH_{\lambda_k}(\bp_{k+1},\br_{k+1})- \cH_{\lambda_k}(\bp_{k},\br_{k})-Q_k\right) \right]) \]
with the Metropolis rule. If the move is accepted, the new state is 
$\tilde{\chi}_{k+1}=(\bp_{k+1},\br_{k+1})$, otherwise we set $\tilde{\chi}_{k+1}=(-\bp_{k},\br_{k})$. The  change of sign for momenta preserves the reversibility of the Markov chain and we take  $W(k\rightarrow k+1) = \cH_{\lambda_{k+1}}(\tilde{\chi}_{k+1})-\cH_{\lambda_{k}}(\tilde{\chi}_{k+1})$.  
A rejection rate of a few percent~\cite{athenes:2006} makes it possible to use a much larger time step, thus saving computational time. Note that when a rejection occurs with probability $1-{\rm P}^{\rm acc}_k$ at iteration $k$, this quantity must be included in the conditional probabilities ${\rm P}_{\rm cond}(\chi_n,n)$ whatever $0 \leq n \leq N$. Because the selecting probability ${\rm P}_{\rm sel}$ considers ratios of conditional probabilities, the Metropolis rejections in the trajectories does not affect the work quantities $W_n$ and the algorithm given in subsection~\ref{rwa} (see ~\cite[Appendix B.3]{athenes:2002b} for the detailed proof). 

Finally, note that the two Ornstein-Uhlenbeck (OU) processes~\eqref{forward0} and~\eqref{forward3} of duration $\Delta t/2$ in the discretization scheme can possibly be merged into a single one of duration  
$\Delta t$. As a result, the discretization (Eqn.~\ref{forward0}-\ref{forward3}), simplifies into a leap-frog scheme~\cite{athenes:2004} that generates a single noise per iteration 
\begin{eqnarray} 
p_{i,k+1/{2}} & = & p_{i,k-1/{2}} + \left( \ell_{i,k} +\tilde{f}_{i,k} \right)\Delta t \hspace{2cm} 
q_{i,k+1} =  q_{i,k} + m_i^{-1} p_{i,k+1/{2}}\Delta t .   \nonumber
\end{eqnarray}
The quantity $\ell_{i,k}\Delta t$ describes the momentum variation during the OU process twice longer
\begin{eqnarray} 
\ell_{i,k}\Delta t=\left(p_{i,k-1/{2}}+\tilde{f}_{i,k} \Delta t/2 \right) \left(e^{-\tilde{\gamma}_i \Delta t}-1\right) + \eta_{i,k} \equiv \left(\ell_{i,k-1/{4}}+\ell_{i,k+1/{4}}\right) \Delta t /2
\nonumber
\end{eqnarray}
where $\eta_{i,k}$  is a normal noise of variance ${(1-e^{-2\tilde{\gamma}_i\Delta t})m_i/\beta}$ while  $\ell_{i,k-1/{4}}\Delta t /2$ and $\ell_{i,k+1/{4}} \Delta t /2$ denote the momentum variations during the OU processes in ~\eqref{forward0} and~\eqref{forward3}. The leap-frog scheme could have been used in combination with the multi-state estimator since here the work does not depend on momenta at integer steps.


\begin{thebibliography}{00} 

\bibitem{torrie:1977} G. Torrie and J. Valleau, Non-physical sampling distributions in Monte Carlo free-energy estimation: Umbrella Sampling, J. Comp. Phys. 23 (1977) 187-199.   
\bibitem{ferrenberg:1989} A. M. Ferrenberg and R. H. Swendsen, Optimized Monte-Carlo Data Analysis, Phys. Rev. Lett. 63 (1989) 1195-1198. 
\bibitem{kumar:1992} S. Kumar, D. Bouzida,R. Swendsen et al., The weighted histogram analysis method for free-energy calculations on biomolecules 1. the method, J. Comp. Chem. 13 (1992) 1011-1021. 
 
\bibitem{chodera:2008} M. Shirts and J. Chodera, Statistically optimal analysis of samples from multiple equilibrium states, J. Chem. Phys. 129 (2008) 124105 (10pp.).  
\bibitem{bennett:1976} C.~H. Bennett, Efficient estimation of free-energy differences from Monte-Carlo data, J. Comp. Phys. 22 (1976) 245-268.  
\bibitem{shirts:2003} M. Shirts, E. Bair, G. Hooker and V. Pande, Equilibrium free energies from nonequilibrium measurements using maximum-likelihood methods, Phys. Rev. Lett. 91, (2003), 140601.

\bibitem{hummer:2001} G. Hummer and A Szabo, Free energy reconstruction from nonequilibrium single-molecule pulling experiments, Proc. Natl. Acad. Sci. U.S.A. 98 (2001) 3658-2010. 
\bibitem{jarzynski:1997} C. Jarzynski, Nonequilibrium equality for free energy differences, Phys. Rev. Lett., 78 (1997), 2690-2693.  
\bibitem{kurchan:1998} J. Kurchan, Fluctuation theorem for stochastic dynamics, J. Phys. A Math. Gen. 31 (1998) 3719-3729. 

\bibitem{Crooks:1998} G. Crooks, Nonequilibrium measurements of free energy differences for microscopically reversible Markovian systems, J. Stat. Phys. 90 (1998) 1481-1487. 

\bibitem{crooks:2000} G. Crooks, Path-ensemble averages in systems driven far from equilibrium, Phys. Rev. E, 61, 2361 (2000) 2361-2366. 

\bibitem{maragliano:2006} L. Maragliano, E. Vanden-Eijnden, A temperature accelerated method for sampling free energy and determining reaction pathways in rare events simulations, Chem. Phys. Lett. 426 (2006) 168-175. 

\bibitem{maragliano:2009}  L. Maragliano, G. Cottone, G. Ciccotti, E. Vanden-Eijnden, Mapping the Network of Pathways of CO Diffusion in Myoglobin, J. Am. Chem. Soc., 132, (2010) 1010–1017. 

\bibitem{abrams:2010} C. Abrams, E. Vanden-Eijnden, Large-scale conformational sampling of proteins using temperature-accelerated molecular dynamics, Proc. Natl. Acad. Sci. U.S.A. 107 (2010) 4961-4966.  

\bibitem{Rosso:2001} L. Rosso, P. Min\'ary, S. Shou, M.E. Tuckerman, On the use of adiabatic molecular dynamics technique in the calculation of free energy profiles, J. Chem. Phys. 116
(2001) 4389-4402.

\bibitem{Rothlisberger:2002} J. Vande Vondele, U. Rothlisberger, Canonical Adiabatic Free Energy Sampling (CAFES): A Novel Method for the Exploration of Free Energy Surfaces, J. Phys. Chem. B 106 (2002) 203-208. 

\bibitem{laio:2002} A. Laio, M. Parrinello, Escaping free-energy minima, Proc. Nat. Acad. Sci. USA 99 (2002) 12562-12566. 

\bibitem{athenes:2002a} M Ath\`enes, Parallel Monte Carlo simulations using a residence weight algorithm, Phys. Rev. E 66 (2002) 016701. 

\bibitem{athenes:2002b} M Ath\`enes, Computation of a chemical potential using a residence weight algorithm, Phys. Rev. E, 66 (2002) 046705. 
\bibitem{athenes:2004} M. Ath\`enes, A path-sampling scheme for computing thermodynamic properties of a many-body system in a generalized ensemble, {Eur. Phys. J. B}, 38 (2004) 651-663. 
\bibitem{adjanor:2005} G. Adjanor  and M. Ath\`enes, Gibbs free-energy estimates from direct path-sampling computations, J. Chem. Phys., 123 (2005) 234104. 

\bibitem{Oberhofer:2005} H. Oberhofer, C. Dellago and P. L. Geissler, Biased Sampling of Nonequilibrium Trajectories: Can Fast Switching Simulations Outperform Conventional Free Energy Calculation Methods? J. Phys. Chem. B, 109 (2005)  6902-6915. 

\bibitem{athenes:2007} M. Ath{\`e}nes, Web ensemble averages for retrieving relevant information from rejected Monte Carlo moves, {Eur. Phys. J. B}, {\bf 58}, 83, (2007). 

\bibitem{LRS2010} T. Leli{\`e}vre, M. Rousset and G. Stoltz, Free-energy computations: a mathematical perspective, Imperial College Press, 2010. 

\bibitem{frenkel:2004} D. Frenkel, Speed-up of Monte Carlo simulations by sampling of rejected states, Proc. Natl. Acad. Sci. U.S.A. {\bf 101}, 17571-17575 (2004) 17571-17575.  

\bibitem{ceperley:1977} D. Ceperley, G. Ghester and M. Kalos, Monte Carlo simulation of a many-fermion system, {Phys. Rev. B} 16 (1977) 3081-3099. 
\bibitem{athenes:2006} G. Adjanor and M. Ath\`enes and F. Calvo, Free energy landscape from path-sampling: application to the structural transition in LJ(38), Eur. Phys. J. B, 53 (2006) 47-60. 
 
\bibitem{athenes:2008} M. Ath\`enes and G. Adjanor, Measurement of nonequilibrium entropy from space-time thermodynamic integration, {J. Chem. Phys.}, 129 (2008) 024116. 

\bibitem{bussi:2009} G. Bussi, T. Zykova-Timan and M. Parrinello, Isothermal-isobaric molecular dynamics using stochastic velocity rescaling, J. Chem. Phys. 130, (2009) 074101. 
\bibitem{chernyak:2006} V. Chernyak and M. Chertkov and C. Jarzynski, Path-integral analysis of fluctuation theorems for general Langevin processes, J. Stat. Mech., (2006) P08001. 
\bibitem{vanden:2009} E. Vanden-Eijnden, Some Recent Techniques for Free Energy Calculations, J. Comp. Chem. 30 (2009) 1737. 

\bibitem{bodineau:2007} T. Bodineau, B. Derrida, Cumulants and large deviations of the current through non-equilibrium steady states, CR Physique, 8 (2007) 540-555.  
\bibitem{crookthesis} G. Crooks, Excursions in Statistical Physics, PhD Thesis, http://threeplusone.com/pubs/GECthesis.  
\bibitem{bochkov77} G N Bochkov  and Yu. E. Kuzovlev, Contribution to general theory of thermal fluctuations in nonlinear-systems, J. Exp. Theor. Phys., 72 (1977) 238-247. 
\bibitem{bochkov81} G N Bochkov and Yu. E Kuzovlev, Non-linear fluctuation-dissipation relations and stochastic-models in non-equilibrium thermodynamics 1 Generalized fluctuation-dissipation theorem, Physica A 106 (1981) 443-479. 
\bibitem{jarzynski:2004} C. Jarzynski, Nonequilibrium work theorem for a system strongly coupled to a thermal environment  {J. Stat. Mech.: Theory Exp.} (2004) P09005. 
\bibitem{jarzynski:2008} J. Horowitz and C. Jarzynski, Comparison of work fluctuation relations, {J. Stat. Mech.: Theory Exp.} P11002, (2007).  
\bibitem{jarzynski:2008b} C. Jarzynski, Nonequilibrium work relations: foundations and applications, Eur. Phys. J. B 59 (2008) 331-340. 

\bibitem{lechner:2007} W. Lechner and C. Dellago, On the efficiency of path sampling methods for the calculation of free energies from non-equilibrium simulations, J. Stat. Mech. 04 (2007) P04001. 

\bibitem{oberhofer:2008} H. Oberhofer and C. Dellago, Optimum bias for fast-switching free energy calculations, Comp. Phys. Com. 179 (2008) 41-45. 
 
\bibitem{dellago:2002} C. Dellago, P. Bolhuis and P. Geissler, Transition Path Sampling, Adv. Chem. Phys. 123 (2002) 1-78. 

\bibitem{stoltz:2007} G. Stoltz, Path sampling with stochastic dynamics : some new algorithms, J. Comp. Phys. 225 (2007) 491-508. 

\bibitem{frenkel:2002} D. Frenkel and B. Smit, Understanding molecular simulation, Academic Press, New York 2002, p389. 

\bibitem{delmas:2007} J.-F. Delmas, B. Jourdain,  Does waste-recycling really improve the multi-proposal Metropolis-Hastings Monte Carlo algorithm?, J. Appl. Probab., 46 (2009), 938-959. 

\bibitem{ackland:2004} G. Ackland, M. Mendelev, D. Srolovitz, S. Han and A. Barashev, Development of an interatomic potential for phosphorous impurities in $\alpha$-Iron, {J. Phys.: Condens. Matter} 16 S2629 (2004). 

\bibitem{bruneval:2007} F. Bruneval, D. Donadio and M. Parrinello, Molecular dynamics study of the solvation of calcium carbonate in water, {J. Phys. Chem. B} {111} (2007) 12219-12227. 
\bibitem{adjanor:2008} G. Adjanor and M. Ath\`enes, Thermodynamic modelling of glasses at atomistic scale, AIP Conference Proceedings, 999 (2008) 186-201. 
%
\bibitem{minh:2008} D. Minh and A. Adib, Optimized free energies from bidirectional single-molecule force spectroscopy, Phys. Rev. Lett. 100, (2008) 180602. 

\bibitem{jarzynski:2006} C. Jarzynski, Rare events and the convergence of exponentially averaged work values, Phys. Rev. E 73 (2006) 046105. 

\bibitem{marinica:2007} M.-C. Marinica and F. Willaime, Orientation of interstitials in clusters in alpha-Fe: A comparison between empirical potentials, Solid State Phen., 129 (2007) 67-74. 

\bibitem{marchese:1986} M. Marchese, G. de Lorenzi, G. Jaccuci and C. Flynn, Jump dynamics and the isotope effect in solid-state diffusion, Phys. Rev. Lett. 57 (1986) 3280-3283. 
\bibitem{Doye:1999} J. Doye  and M. Miller and D. Wales, The double-funnel energy landscape of the 38-atom Lennard-Jones cluster, J. Chem. Phys. 110 (1999) 6896-6906. 

\bibitem{CalvoNFD00} F. Calvo and J. P. Neirotti and D. L. Freeman and J. D. Doll, Phase changes in 38-atom Lennard-Jones clusters, J. Chem. Phys., 112 (2000) 10350-10357. 

\bibitem{Bogdan:2006} D. Wales and Tetyana Bogdan, Potential energy and free energy landscapes, J. Phys. Chem. B 110 (2006) 20765-20776. 

\bibitem{Steinhardt:1983} P. Steinhard, D. Nelson and M. Ronchetti, Bond Orientational order in Liquids and Glasses, {Phys. Rev. B}, 28 (1983) 784-805. 
\bibitem{Wales:2003} D. Wales, Energy landscapes, Cambridge University Press (2003). 
\bibitem{wang:2001} F. Wang and D. P. Landau, Efficient multiple-range random walk algorithm to calculate density of states, Phys. Rev. Lett. 86 (2001) 2050-2053. 
 
\bibitem{calvo:2002} F. Calvo, Sampling along reaction coordinates with the Wang-Landau method, Mol. Phys. 100 (2002) 3421-3427.
\bibitem{coluzza:2005} I. Coluzza and D. Frenkel, Virtual-move Parallel Tempering {ChemPhysChem}, 6 (2005) 1779-1783. 

\bibitem{lelievre:2007} T. Leli\`evre, M. Rousset and G. Stoltz, Computation of free energy profiles with parallel adaptive dynamics, J. Chem. Phys. 126 (2007) 134111. 
\bibitem{barducci:2008} A. Barducci, G. Bussi and M. Parrinello {Phys. Rev. Lett.}, Well-tempered metadynamics: A smoothly converging and tunable free-energy method, 100 (2008) 020603. 

\bibitem{frenkel:2005} D. Frenkel, Waste-recycling Monte Carlo, in ``Computer Simulations in Condensed Matter Systems'', Lect. Notes Phys. 703 (2006) 127.  

\bibitem{athenes:2008c} M. Ath\`enes and F. Calvo, Multiple-Replica Exchange with Information Retrieval, ChemPhysChem 9 (2008) 2332-2339. 

\bibitem{esselink:1995} K. Esselink, L. Loyens, B. Smit, Parallel Monte Carlo Simulations, Phys. Rev. E 51 (1992) 1560-1568. 

\bibitem{lanore:1974} J.-M. Lanore, Simulation of the evolution of defects in a lattice by the Monte Carlo method, Radiation Effects,~22 (1974) 153-162. 
\bibitem{bortz:1975} New algorithm for Monte-Carlo Simulation of Ising Spin systems, A.~B. Bortz, M.~H. Kalos, and J.~L. Lebowitz, {{J. Comp. Phys.}} {~17} (1975) 10-18. 
\bibitem{athenes:1997} M.~Ath\`enes, P.~Bellon, and G.~Martin, Identification of novel diffusion cycles in B2 ordered phases by Monte Carlo simulation, Phil. Mag. A, {76} (1997) 565-585. 
\bibitem{mason:2004} D.~R. Mason, R.~E. Rudd, and A.~P. Sutton, Stochastic kinetic Monte Carlo algorithms for long-range Hamiltonians, Comput. Phys.Commun. 160 (2004)  140-157. 
\bibitem{bulatov:2006} T. Oppelstrup, V. Bulatov, G. Gilmer, M. Kalos, and B. Sadigh, First-Passage Monte Carlo Algorithm: Diffusion without all the hops. Phys. Rev. Lett. 97 (2006) 230602. 

\bibitem{wales:2009} D. Wales, Calculating rate constants and committor probabilities for transition networks by graph transformation, J. Chem. Phys.~130 (2009) 204111. 

\bibitem{boulougouris:2005} G. Boulougouris and D. Frenkel, Monte Carlo sampling of a Markov web, J. Chem. Theory Comp. 1 (2005) 389-393. 

\end{thebibliography}
\end{document}